\documentclass[journal]{IEEEtran}

\usepackage{amssymb, graphicx, cite, amsmath, psfrag, minibox, mathrsfs, bigints, stfloats, color, upgreek, gensymb, epstopdf, textcomp, amsthm, amsfonts, euscript, calligra}

\usepackage[font=footnotesize]{subcaption}
\usepackage[font=footnotesize]{caption}

\usepackage{mathtools, amsfonts, multirow, array}
\usepackage{verbatim, mdwtab, relsize, cleveref, mdwmath}
\usepackage{algorithmicx}
\usepackage{algpseudocode}
\usepackage[ruled]{algorithm2e}

\usepackage{setspace}
\usepackage{float}

\usepackage{tikz}
\usetikzlibrary{shapes,arrows}
\usepackage{textcomp}

\makeatletter
\newcommand{\leqnomode}{\tagsleft@true}
\newcommand{\reqnomode}{\tagsleft@false}
\makeatother

\def\pioh{\widehat{\pi}_0}
\def\pilh{\widehat{\pi}_1}
\def\P0{P^{(0)}}

\def\Pbar{\overline{P}_{\text{av}}}
\def\Ibar{\overline{I}_{\text{av}}}

\def\SUTx{SU$_{\text{tx}}$}
\def\SURx{SU$_{\text{rx}}$}

\ifCLASSINFOpdf
\else
\fi

\begin{document}
\title{Beam Selection and Discrete Power Allocation in Opportunistic Cognitive Radio Systems with Limited Feedback Using ESPAR Antennas}

\author{Hassan Yazdani,
	  Azadeh Vosoughi,~\IEEEmembership{Senior Member,~IEEE},
       Xun Gong,~\IEEEmembership{Senior Member,~IEEE}\\
        University of Central Florida\\
        E-mail: {\tt \normalsize h.yazdani@knights.ucf.edu, azadeh@ucf.edu, xun.gong@ucf.edu} 
        
        \thanks{Parts of this research were presented at {\it 53nd Annual Conference on Information Sciences and Systems (CISS)}, 2019 \cite{CISS2019}. This research is supported by the NSF under grant ECCS-1443942.}        
}
%
%
%
\markboth{}%
{Shell \MakeLowercase{\textit{et al.}}: Bare Demo of IEEEtran.cls for Journals}
%
%
\newcounter{MYeqcounter}
\multlinegap 0.0pt                     
%
%
\maketitle
%
%
\begin{abstract}
We consider an opportunistic cognitive radio (CR) system consisting of a primary user (PU), secondary transmitter (\SUTx), and secondary receiver (\SURx), where \SUTx ~is equipped with an electrically steerable parasitic array radiator (ESPAR) antenna with  {\color{blue} beam steering capability} 
for sensing and communication, and there is a limited feedback channel from \SURx ~to \SUTx. Taking a holistic approach, we develop a framework for integrated sector-based spectrum sensing and sector-based data communication. Upon sensing the channel busy, \SUTx ~determines the beam corresponding to PU's orientation. Upon sensing the channel idle, \SUTx ~transmits data to \SURx, using the selected beam corresponding to the strongest channel between \SUTx ~and \SURx.  
We formulate a constrained optimization problem, where \SUTx-\SURx ~link ergodic capacity is maximized, subject to average transmit power and interference constraints, and the optimization variables are sensing duration, thresholds of channel quantizer at \SURx, and transmit power levels at \SUTx. Since this problem is non-convex we develop a suboptimal computationally efficient iterative algorithm to find the solution. Our {\color{blue} numerical}  results {\color{blue} quantify the capacity improvement provided by the ESPAR antenna 
and } demonstrate that our CR system yields 
lower outage and symbol error probabilities, compared with a CR system that its \SUTx ~has an omni-directional antenna.
\end{abstract}
%
%
\vspace{-1mm}
%
\IEEEpeerreviewmaketitle
\vspace{-1mm}
\section{Introduction}\label{Se1}
%
%
%
%
\vspace{-1mm}
\subsection{{\color{blue}Overview} and Background}
\vspace{-1mm}
Cognitive radio (CR) is a promising solution that enhances spectrum utilization by allowing an unlicensed or secondary user (SU) to access licensed bands in a such way that its imposed interference on license holder primary users (PUs) is limited, and hence fills the spectrum holes in time and/or frequency domains \cite{Arslan, Nalla, Liang2, Gursoy1, Beaulieu, Gursoy2}. There is a rich literature on underlay CR systems, where PUs and SUs are allowed to transmit simultaneously and in the same frequency band, as long as the interference caused by SUs to PUs stays below a pre-determined threshold. While underlay systems do not require spectrum sensing to detect PU's activities, they demand coordination between PUs and SUs (to obtain channel state information (CSI) of PU links at SUs) that is not always feasible. On the other hand, interweave or opportunistic CR systems utilize spectrum sensing to enable SUs to use a licensed frequency band during a time interval, only if PUs are not using that frequency band within that time interval, implying that coordination between PUs and SUs to acquire CSI is not needed.
\par Reconfigurable antennas (RA) \cite{Mahmoud1, Mahmoud2}, with the capabilities of dynamically modifying their characteristics (e.g., operating frequency, beamwidth, radiation pattern, polarization) can improve the spectral efficiency (well beyond what is attainable with omni-directional antennas), via beam steering and utilizing the spectrum white spaces in spatial (angular) domain.  RAs have been used to design directional wireless and millimeter-wave communication systems and surveillance \cite{Jafarkhani}. An electrically steerable parasitic array radiator (ESPAR) antenna is a special kind of RAs, that has been used for identifying the spectral holes in spatial domain in CR systems. ESPAR divides the angular domain into several sectors (beams) and switches between beampatterns of sectors in a time-division fashion (only one of $M$ beams is active at a time). The ESPAR antenna relies on a single RF front end (an active element) coupled to several passive or parasitic elements (mutually coupled to the active one) to steer beams in prescribed directions \cite{MILCOM, GongESPAR1}. The active element is connected to the transmitter/receiver circuit and the parasitic elements are reactively loaded. Since only one RF chain is needed, the power consumption, cost, and hardware complexity are significantly reduced. The mutual coupling between the ESPAR antenna elements is created by reducing the spacing between them, which makes this antenna suitable for small mobile devices. For CR systems, the ESPAR antennas provide an improved spectrum sensing, due to a signal-to-noise ratio (SNR) increase for transmission and reception of directional signals, and limit out-of-band interference to and from PUs \cite{Spatial_SS_Parasitic}. The ESPAR antennas have the capability of transmitting multiple data streams by signal projection on beamspace basis \cite{Random_Aerial}. Also, they can be used for blind interference alignment through beampattern switching \cite{BlindInterference}. ESPAR antennas have been used in \cite{Clerckx1}, to provide an end-to-end solution for  practically implementable cloud radio access networks. RAs can enhance performance of multiple-input multiple-output (MIMO) systems, via enabling joint beam and antenna selection optimization \cite{Ghrayeb2, Ghrayeb1, Evans}. 
\par Motivated by the benefits of ESPAR antennas, in this work we consider an opportunistic CR system, where SU transmitter (\SUTx) is equipped with an ESPAR antenna, and \SUTx ~uses the directions (identified during spectrum sensing phase),
for data communication with SU receiver (\SURx) with an optimized discrete power level. To the best of our knowledge, this is the first work that proposes a holistic system design for {\it integrated} sector-based spectrum sensing and sector-based data communication for opportunistic CR systems using ESPAR. 
%
%
%
\vspace{-2mm}
\subsection{Spectrum Sensing in Opportunistic CR Systems Using ESPAR Antennas}\label{Section1B}
\vspace{-1mm}
Considering the ESPAR antennas, the authors in \cite{Spatial_SS_Parasitic, Max_min_SSS_Parasitic, Directional_SS_ESPAR, Blind_Spatial_SS} designed detectors, based on the received signal energy in different beams, and also eigenvalue-based 
detectors, via constructing the covariance matrix that captures the signal correlation across beams. The advantages of spectrum sensing using ESPAR antennas are twofold. First, the SNR gain from the directional beampatterns increases the probability of detecting PU's activities within that beam, and hence decreases the chance of causing interference on PU. Second, the discovered unoccupied beams in spatial (angular) domain during spectrum sensing represents directional transmit/receive opportunities for SUs, which can be utilized to increase spectral efficiency (opportunities that would be missed when using an omni-directional antenna at \SUTx).
%
%
%
\vspace{-2mm}
\subsection{Beam Selection for Data Communication in Underlay CR Systems Using RA Antennas}\label{Section1C}
\vspace{-1mm}
Selecting the best beam for data communication has been considered before in \cite{RUB_Selection_Yang, Alouini3, Alouini4, Liang1} using general/traditional directional antennas (with multi-RF chain) and in \cite{Ghrayeb2, Ghrayeb1, Evans, SNR_Reconfig, ArslanESPAR} using RAs for underlay CR systems. For instance, the authors in \cite{Ghrayeb2} considered space-shift keying (SSK) signaling and investigated the best beam selection method that improves SSK performance (in terms of throughput, system complexity, error probability), via minimizing the Rician $K$-factor and the correlation coefficient between the antenna beams. Considering SSK signaling, the authors in \cite{Ghrayeb1} proposed a beam selection scheme that improves the performance of the secondary system, while meeting the transmit power and outage interference constraints (without any feedback from the receiver). 
For a multiuser orthogonal frequency-division multiple access (OFDMA) underlay CR network, the authors in \cite{ArslanESPAR} developed a game theoretical framework for joint optimization of beam and subcarrier selection at each SU, such that the overall network capacity is maximized, while the interference constraint on the primary network is met. The authors in \cite{Evans} studied a different scenario where \SURx ~is equipped with multiple RAs and beam selection is conducted  based on the channel between \SUTx ~and \SURx ~and considering  several performance metrics (achievable rate, error and outage probabilities). 
The work in \cite{SNR_Reconfig} shows that comparing RA and  traditional antenna selection, the former can offer significant improvements in SNR.
%
%
%
\vspace{-2mm}
\subsection{Beamforming for Data Communication Using ESPAR Antennas}\label{Section1D}
\vspace{-1mm}
\par A related research thrust in the context of ESPAR antennas for CR systems is designing adaptive beampatterns (also called beamforming) \cite{Spatial_SS_Parasitic , Pow_ESPAR, Ohira, MVDR, RobustBeamForming}. 
For instance, the authors in \cite{Spatial_SS_Parasitic} proposed an adaptive beamforming algorithm, that numerically optimizes the beampattern and antenna efficiency, and creates beampattern nulls to protect PU from unwarranted transmissions by SUs.
However, such a design approach from a mathematical perspective is very challenging, due to the tunable reactive loads, which renders the problem a non-convex optimization without any closed form solution. Furthermore, implementing the design incurs high computational complexity. \cite{Pow_ESPAR} utilized the switchable weakly-correlated patterns of an ESPAR antenna for underlay CR systems, in order to maximize the transmitted power to \SURx ~while limiting interference imposed on PU. \cite{Ohira} presented several numerical methods for SNR optimization and beam-null steering via maximizing the cross-correlation coefficient between a known reference (pilot) and the received signals. \cite{MVDR} proposed a minimum variance distortionless response (MVDR) beamforming method which steers the beam at the desired direction and places nulls at the interfering directions. 
 \cite{RobustBeamForming} designed a robust precoding scheme for a MIMO CR system, where the CR interference channel is completely unknown. 
 {\color{blue} We note that the implicit assumption in these works is that the directions of PU and \SURx ~with respect to \SUTx ~are known and hence the shape of the beam can be optimized accordingly.  Thus, the beamforming designs in \cite{Spatial_SS_Parasitic , Pow_ESPAR, Ohira, MVDR, RobustBeamForming} are not directly applicable to a problem 
 where these direction information are missing.}
 %
 %
\vspace{-1mm}
{\color{blue}
\subsection{Knowledge Gap and Our Motivation }\label{Section1E}
\vspace{-0.5mm}
Our review of the literature indicates that the studies on optimizing spectrum sensing and optimizing data communication have been pursued as two separate research thrusts: the works cited in Section \ref{Section1B} focus on spectrum sensing in opportunistic CR systems, whereas the works in Section \ref{Section1C} and  Section \ref{Section1D} focus on data communication in underlay CR systems. 
The developed beam selection and beamforming schemes in these works are specifically tailored for underlay CR systems, which do not require spectrum sensing to detect PU's activities, and rely on the knowledge obtained from coordination between PUs and SUs. Evidently, the literature lacks a holistic system design, that integrates spectrum sensing and data communication in a cohesive manner for opportunistic CR systems. Such a holistic system design needs to take into consideration the effect of imperfect spectrum sensing on data communication optimization. This is the motivation behind our work. 
Taking full advantage of beam steering capability of the ESPAR antenna (the capability of choosing one beam among $M$ beams),
we propose an integrated design for an opportunistic CR system, where \SUTx ~is equipped with an ESPAR antenna. We leverage on the beam steering capability of the ESPAR antenna for both spectrum sensing and data communication optimization. Different from the state-of-the-art, our proposed integrated design incorporates induced errors due to: (i)  imperfect spectrum sensing and determining the correct beam corresponding to PU's location, such errors affect the interference imposed on PU; (ii) selecting the best beam for data communication over \SUTx-\SURx ~link. }
%
%
%
\vspace{-1mm}
\subsection{Our Contributions and Paper Organization}
%
%
In this paper, we consider an opportunistic CR system consisting of a PU, \SUTx, and \SURx, where \SUTx ~is equipped with an ESPAR antenna with the capability of choosing one sector among $M$ sectors for its data transmission to \SURx. During the initial {\it channel sensing phase} \SUTx ~senses the channel and monitors the activity of PU. While being in this phase, \SUTx ~determines the beam corresponding to the location (orientation) of PU based on the received signal energy. \SUTx ~stays in this phase as long as the channel is sensed busy. It leaves this phase and enters {transmission phase} when the channel is sensed idle. The transmission phase itself consists of two phases: {\it{channel training phase}} followed by {\it{data transmission phase}}. During the former phase, \SUTx ~sends pilot symbols to enable channel training and estimation at \SURx ~as well as selection of the strongest channel among all beams between \SUTx-\SURx ~for data transmission. Also, \SURx ~employs an $n_b$-bit quantizer to quantize the gain of the selected beam. Then, \SURx ~feeds back the index of  the selected  beam as well as the $n_b$-bit  representation of the index of the quantization interval over an error-free bandwidth limited feedback link to \SUTx, so \SUTx ~can optimally adapt its discrete power level accordingly. The main contributions of this paper can be summarized as follows: 
\begin{itemize}
\item  Given this system model, we formulate a novel optimization problem, aiming at maximizing the constrained ergodic capacity of \SUTx-\SURx ~link, subject to average interference and average transmit power constraints.
\item Our problem formulation takes into consideration the effect of imperfect spectrum sensing as well as the error due to incorrect determination of the beam corresponding to PU's location (and its corresponding effect on imposed average interference) occurred during {\it channel sensing and monitoring phase}. 
\item  Our problem formulation also takes into account the probability of correct determination of the strongest beam for data transmission from \SUTx ~to \SURx, occurred during {\it channel training phase}. It also incorporates the impact of CSI quantization on the constrained optimization problem in hand. 
\item  We solve the formulated problem and optimize the duration of spectrum sensing, thresholds of CSI quantizer, and discrete transmit power levels (to be employed at \SUTx) corresponding to CSI quantization intervals. 
\item  For our system model, we provide closed form expressions for outage and symbol error probabilities. 
\item   To the best of our knowledge, this is the first work that adopts a holistic approach to design an opportunistic CR system using ESPAR antennas and integrates sector-based spectrum sensing and sector-based data communication. All cited works use RAs for enhanced communication in underlay CR systems. Utilizing ESPAR antennas for opportunistic spectrum sharing systems is a highly promising solution to enhance the performance of secondary links, while satisfying the constraints set by the primary links \cite{ArslanESPAR}.
\item  This work is different from our preliminary works in \cite{ICASSPpaper, AsilomarPaper, GlobalSIP3}, where we have considered a simpler constrained optimization problem (with different optimization variables, including continuous transmit power and direction of antenna steering of \SUTx), assuming that \SUTx ~knows the direction (angle) corresponding to PU's activities. It is also different from our work in \cite{CISS2019}, where we have assumed that \SUTx ~knows the location of \SURx ~and PU, and, perfect CSI of \SUTx-\SURx ~link is available and used for transmit power adaptation. This work is different from \cite{Cabric}, in which the direction of PU is estimated at \SUTx ~(i.e., it is not based on determining the sector).  
\item Taking advantage of the additional degrees of freedom offered by ESPAR antennas with variable beam directions, we improve the spectral efficiency and reduce implementation complexity of opportunistic spectrum sharing systems, while fulfilling an average interference constraint imposed on PU. {\color{blue} Our simulations demonstrate and quantify the capacity improvement provided by the ESPAR antenna, in terms of average transmit power $\Pbar$ and average interference  $\Ibar$ constraints. For instance, at $\Pbar\!=\!12$\,dB, $\Ibar\!= \!-6$\,dB, the capacity of our CR system is  $1.83$  times larger than the capacity of a CR system that its \SUTx ~has an omni-directional antenna.}
\item Our numerical results show that with only a small number of feedback bits the capacity of our opportunistic CR system approaches to its baseline, which assumes the full knowledge of unquantized \SUTx-\SURx ~channel gain at \SUTx.
\end{itemize}
%
%
%
%
\par The remainder of the paper is organized as follows. Section \ref{Se2} explains our system model and problem statement. Section \ref{Sect3} characterizes the objective function and the constraints of our optimization problem, in terms of the optimization variables. Specially, in Section \ref{GLRT_SS} we describe our binary energy-based detector for detecting PU activity and in Section \ref{PU_Beam_Selection} we express how \SUTx ~determines the beam corresponding to PU. In Section \ref{SubSe3} we discuss how \SURx ~determines the strongest channel between \SUTx-\SURx ~and we obtain the probability of selecting the true strongest channel among all beams. The problem is formalized and solved in Section \ref{Sect4} and the closed form expressions for outage  and symbol error probabilities are given in Section \ref{Sect5}. Section \ref{SimResults} presents our simulation results and Section \ref{Conclusion} concludes the paper. 
%
%
\section{System Model and Problem Statement}\label{Se2}
%
%
%
%
\subsection{Background on ESPAR Antennas}\label{SysModelA}
%
%
\par
\begin{figure}[!t]
\vspace{-0mm}
\centering
	\begin{subfigure}[b]{0.35\textwidth}  
		\centering	              
		\psfrag{o1}[Bl][Bl][1]{$\cdot$}
		\psfrag{o2}[Bl][Bl][1]{$\cdot$}
		\psfrag{o3}[Bl][Bl][1]{$\cdot$}
		\psfrag{jx1}[Bl][Bl][0.58]{\!\!$j x_1$}
		\psfrag{jx2}[Bl][Bl][0.58]{\!\!$j x_2$}
		\psfrag{jx3}[Bl][Bl][0.58]{\!\!$j x_3$}
		\psfrag{jx4}[Bl][Bl][0.58]{$j x_4$}
		\psfrag{jx5}[Bl][Bl][0.58]{$j x_5$}
		\psfrag{jxM}[Bl][Bl][0.58]{$j x_M$}
		\psfrag{2M}[Bl][Bl][0.58]{$\frac{2 \pi}{M}$}
		\psfrag{a/M}[Bl][Bl][0.58]{$\frac{2 \pi}{M}$}
		\psfrag{3a/M}[Bl][Bl][0.58]{$\frac{2 \pi}{M}$}
		\psfrag{#1}[Bl][Bl][0.58]{$\#1$}
		\psfrag{#2}[Bl][Bl][0.58]{$\#2$}
		\psfrag{#3}[Bl][Bl][0.58]{$\#3$}
		\psfrag{#4}[Bl][Bl][0.58]{$\#4$}
		\psfrag{#0}[Bl][Bl][0.58]{$\#0$}
		\psfrag{#5}[Bl][Bl][0.58]{$\#5$}
		\psfrag{#M}[Bl][Bl][0.58]{$\#M$}
		\psfrag{Vs}[Bl][Bl][0.58]{$V_s$}
		\psfrag{r}[Bl][Bl][0.58]{$r$}
		\psfrag{source}[Bl][Bl][0.58]{\!source}
		\includegraphics[width=60mm]{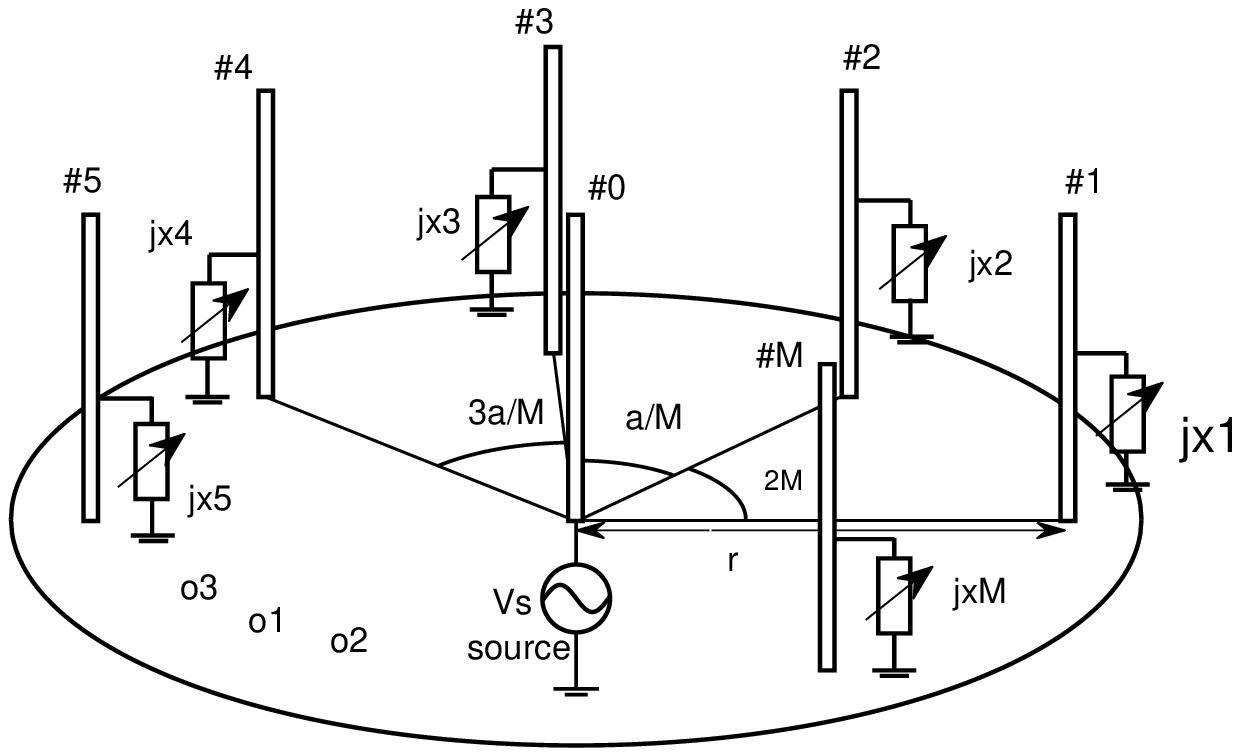}
		\caption{} 
		\vspace{2mm}
		\label{ESPAR_figure}          	         
	\end{subfigure}
     \vspace{0mm}
     \begin{subfigure}[b]{0.35\textwidth}
		\centering
		\psfrag{k1}[Bl][Bl][0.58]{\!$\kappa_1$}
		\psfrag{k2}[Bl][Bl][0.58]{\!$\kappa_2$}
		\psfrag{k3}[Bl][Bl][0.58]{\!$\kappa_3$}
		\psfrag{k4}[Bl][Bl][0.58]{\!\!$\kappa_4$}
		\psfrag{k5}[Bl][Bl][0.58]{\!\!$\kappa_5$}
		\psfrag{k6}[Bl][Bl][0.58]{$\kappa_6$}
		\psfrag{k7}[Bl][Bl][0.58]{\!$\kappa_7$}
		\psfrag{k8}[Bl][Bl][0.58]{$\kappa_8$}
		\includegraphics[width=34mm]{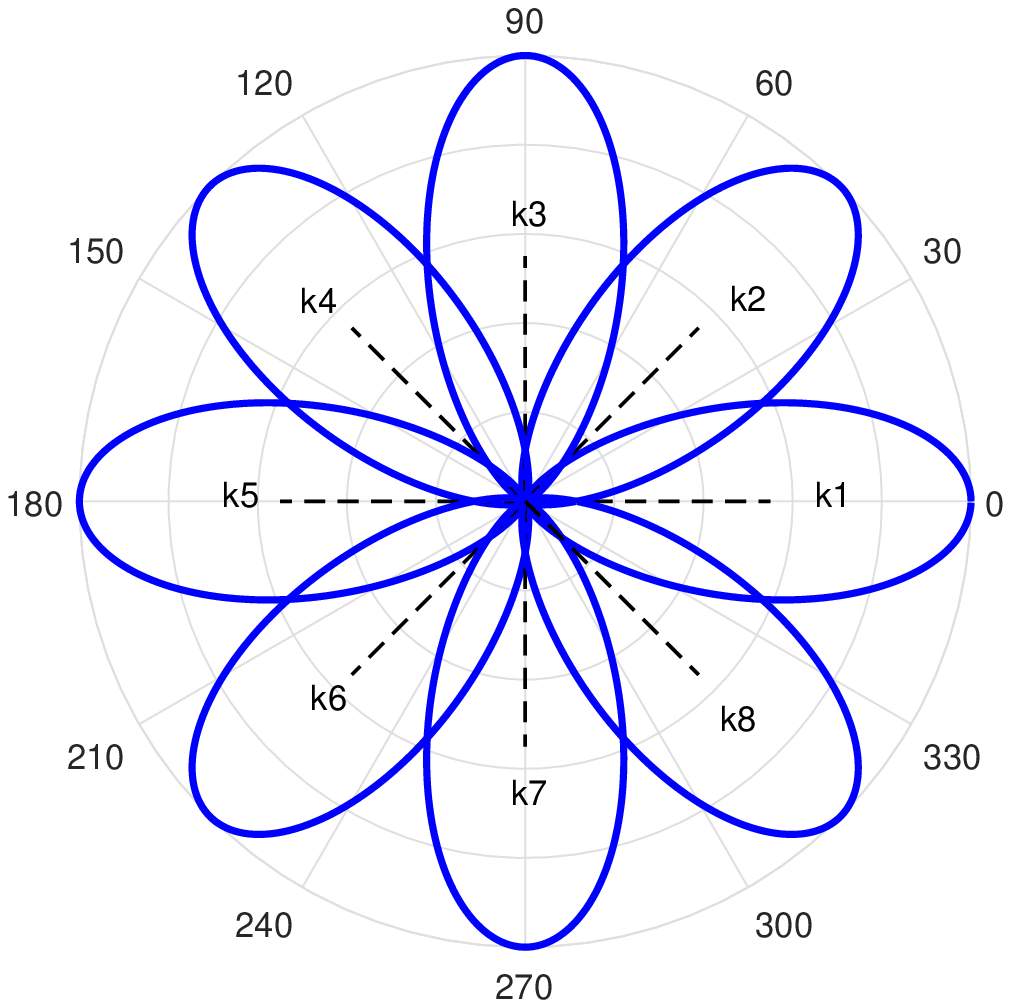}
		\caption{} 
		\label{Pattern_fig}           	  
     \end{subfigure}%
     \vspace{-1mm}
\caption{The ESPAR antenna structure and its beampatterns, (a) The ESPAR antenna structure, (b) Beampatterns  of an ESPAR with $8$ parasitic elements, assuming the Gaussian radiation pattern in \eqref{pattern_formula}.}
\vspace{-4mm}
\end{figure}

 The ESPAR antenna is a circular array, comprised of one active element and $M$ parasitic elements symmetrically surrounding the active element, and the radius of the array is $r < \lambda_c/2$, where  $\lambda_c$ is the carrier wavelength \cite{MILCOM}. Fig. \ref{ESPAR_figure} depicts an ESPAR structure. The active element is connected to the single RF chain, while $M$ parasitic elements (which are mutually coupled to the active element) are short-circuited and loaded by $M$ variable reactive loads. Let $x_m$ be the reactive load of $m$-th element and vector $\boldsymbol{x} = [x_1, \ldots, x_M ]$ denote the reactive loads of all $M$ parasitic elements. By adjusting these reactive loads, the beampatterns of the ESPAR antenna are designed such that the angular space is divided into $M$ spatial sectors or beams\footnote{Throughout this paper, ``sector'' and ``beam'' are used interchangeably.}. In particular, to design the beampattern corresponding to the first beam, entries of vector $\boldsymbol{x}_1$ are optimized such that the beam gain is maximized at an angle (for example angle $0^\degree$)  \cite{MILCOM}. Since the ESPAR antenna structure is symmetric, the beampattern corresponding to the second beam can be obtained by circularly shifting the entries of $\boldsymbol{x}_1$, such that the beam gain is maximized at angle $\kappa_2= \frac{2 \pi}{M}$. Repeating this $M$ times one can obtain $M$  beampatterns corresponding to $M$ beams such that the beampattern corresponding to the $m$-th beam achieves its maximum at angle $\kappa_m= \frac{2 \pi(m-1)}{M}$ for $m=1, \ldots, M$. 
It is noteworthy that the ESPAR antenna can provide an omni-directional beampattern if the reactive loads of all parasitic elements are chosen equal (omni-directional mode).  
%
%
%

%
%
%
\par Similar to \cite{Cabric}, to mathematically model the radiation pattern (antenna pattern) of the ESPAR antenna, we adopt the Gaussian pattern  in $x\!-\!y$ azimuth plane in terms of angle $\phi$ given by
\vspace{-1mm}
\begin{equation}\label{pattern_formula}
p(\phi) = A_1 + A_0 ~ e^{ -B \left ( \frac{ \mathcal{M(\phi)}} {\phi_{\text{3dB}}} \right )^2 }, 
\vspace{-1mm}
\end{equation}
%
%
\vspace{-3mm}
\begin{equation}
\mathcal{M(\phi)} = \text{mod}_{2 \pi} ( \phi + \pi ) - \pi, 
\vspace{-1mm}
\end{equation}
%
$\text{mod}_{2 \pi} ( \phi  )$ denotes the remainder of $\frac{\phi}{2 \pi}$, constant $B=\ln(2)$, $\phi_{3\text{dB}}$ is the 3-dB beamwidth, {$A_1$} and $A_0$ are two constant antenna parameters. 
The radiation pattern of $m$-th sector at angle $\phi$ is 
\vspace{-1mm}
\begin{equation}\label{pm}
p_m(\phi ) = p(\phi - \kappa_m)  ~~~~\text{for} ~m=1, \ldots, M. 
\vspace{-1mm}
\end{equation}
%
In Fig. \ref{Pattern_fig}, the beampatterns of an ESPAR antenna with $8$ parasitic elements are shown. In this paper, we discuss the received or transmitted signal at $m$-th sector of \SUTx. This means that, during the signal reception or transmission, the reactive loads of all $M$ parasitic elements (i.e., the entries of vector $\boldsymbol{x}$) are set and tuned  such that the beampattern corresponding to the $m$-th beam is generated. Note that in our work we assume the reactive loads (i.e., the entries of vector $\boldsymbol{x}$ and thus the shapes of beampatterns or equivalently the radiation patterns of $M$ sectors) are determined by the ESPAR antenna designer. Given the antenna design, we focus on how the sector-based structure of this ESPAR antenna can be exploited to enhance the system performance of our opportunistic CR system, in which \SUTx ~optimizes its sector-based data communication to \SURx ~according to the results of its sector-based channel sensing.
%
%
\vspace{-2mm}
\subsection{Geometry of Our Opportunistic CR System}\label{Section2B}
\vspace{-1mm}
\par
\begin{figure}[!t]
\centering
\psfrag{g}[Bl][Bl][0.65]{$g$}
\psfrag{gss}[Bl][Bl][0.65]{$g_\text{ss}$}
\psfrag{gsp}[Bl][Bl][0.65]{$g_\text{sp}$}
\psfrag{SUtx}[Bl][Bl][0.7]{\SUTx}
\psfrag{SUrx}[Bl][Bl][0.7]{\SURx}
\psfrag{PU}[Bl][Bl][0.7]{PU}
\psfrag{phiSR}[Bl][Bl][0.65]{$\phi_\text{SR}$}
\psfrag{phiPU}[Bl][Bl][0.65]{$\phi_\text{PU}$}
\includegraphics[width=40mm]{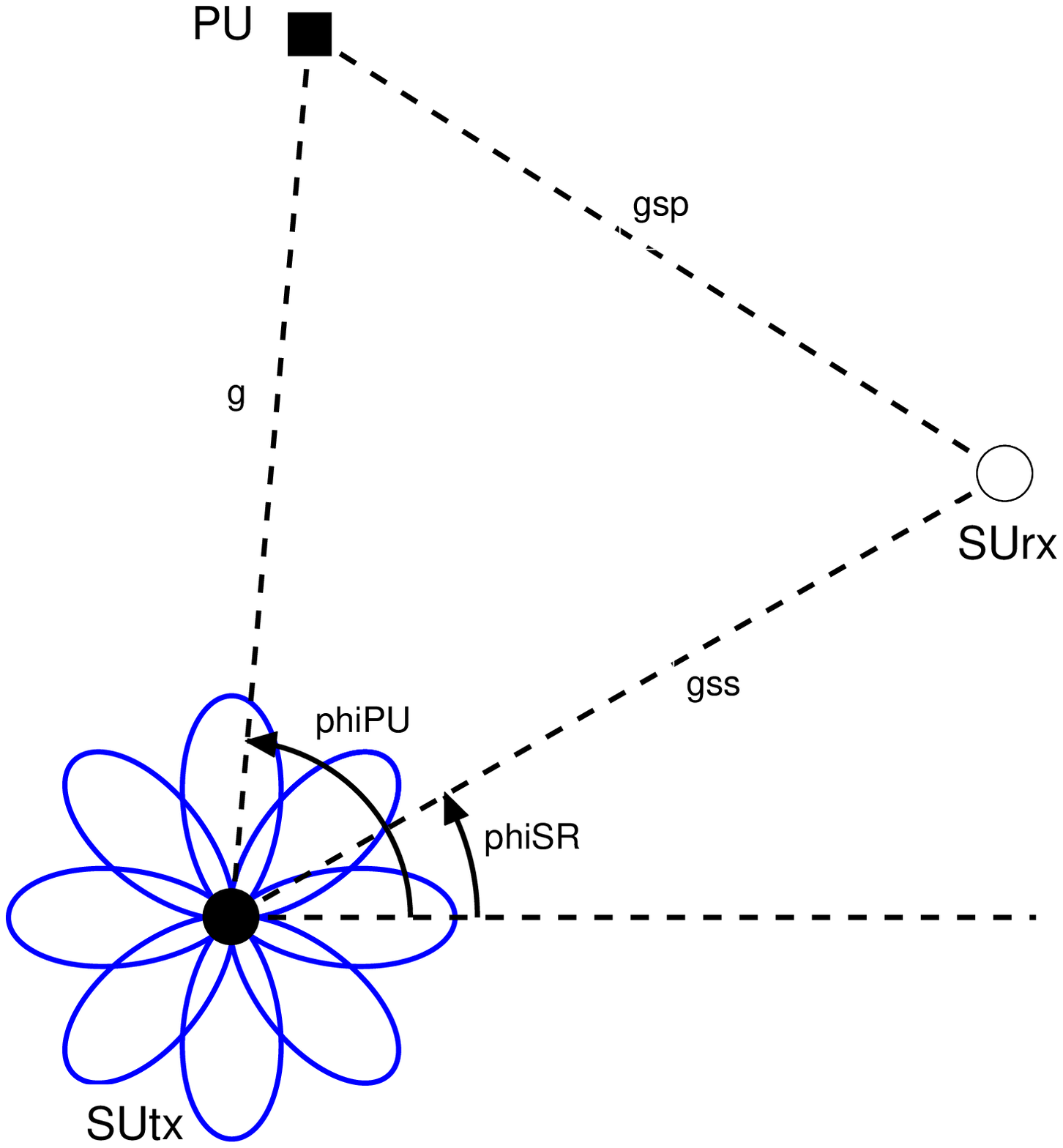}
\vspace{0mm}  
\caption{Our CR system with an ($M\!+\!1$)-element ESPAR antenna at \SUTx ~and omni-directional antennas at \SURx ~and PU. } 
\label{SystemModelFig}      
\vspace{-1mm}  
\end{figure}

 Our CR system model is illustrated in Fig. \ref{SystemModelFig}, consisting of a PU and a pair of \SUTx ~and \SURx. We note that PU in our system model can be a primary transmitter or receiver. We assume when PU is active it is engaged in a bidirectional communication with another PU, which is located far from \SUTx ~and hence its activity does not impact our analysis.
We assume \SUTx ~is equipped with an ($M\!+\!1$)-element ESPAR antenna (for channel sensing and communication) with the capability of choosing one sector among $M$ sectors for its data transmission to \SURx, while \SURx ~and PU use omni-directional  antennas. The reason for this assumption is to focus on quantifying the capacity improvement provided by the ESPAR antenna at \SUTx, in the presence of channel sensing error as well as  average transmit power constraint 
and average interference constraint. 
We also assume there is an error-free bandwidth limited feedback channel from \SURx ~to \SUTx ~(where the channel bandwidth is measured in terms of the number of bits sent over the channel \cite{Alouini, Evans_WSN}, to help \SUTx ~select the best sector for its data transmission to \SURx ~and also to provide \SUTx ~with the quantized channel gain of the selected beam, so \SUTx ~can adapt its discrete power level accordingly. The direction (orientation) of PU and \SURx ~with respect to \SUTx ~are denoted by angles $\phi_\text{PU}$,  and $\phi_\text{SR}$, receptively. Clearly, in our problem \SUTx ~does not know these directions  or angles  (otherwise, the beam selection at \SUTx ~for data transmission would become trivial).
\par Let $h$, $h_\text{ss}$, $h_\text{sp}$ denote the fading coefficients of channels between \SUTx ~and PU, \SUTx ~and \SURx, and \SURx ~and PU, respectively, when the ESPAR antenna of \SUTx ~is in omni-directional mode. We model these fading coefficients as independent circularly symmetric complex Gaussian random variables. We assume $g=|h|^2$, $g_\text{ss}=|h_\text{ss}|^2$ and $g_\text{sp}=|h_\text{sp}|^2$ are independent  exponentially distributed random variables  with mean $\gamma$, $\gamma_\text{ss}$ and $\gamma_\text{sp}$, respectively\footnote{{\color{blue} We note that the distances between users are included in the small scale fading model \cite{GoldSmith}. In particular, we assume that the mean values are $\gamma \!=\! (d_0/d)^\epsilon, \gamma_\text{ss}\!=\!(d_0/d_\text{ss})^\epsilon, \gamma_\text{sp}\! =\!(d_0/d_\text{sp})^\epsilon $, where $d_0$ is the reference distance, $\epsilon$ is the path-loss exponent, and $d, d_\text{ss}$ and $d_\text{sp}$ are the distances between \SUTx ~and PU, \SUTx ~and \SURx, and \SURx ~and PU, respectively. }}
Since in our problem SUs and PU cannot cooperate, SUs cannot estimate $g$ and $g_\text{sp}$. However, we assume that \SUTx ~knows the channel statistics, i.e., the mean values $\gamma$ and $\gamma_\text{sp}$. Let $\psi_m$ and $\chi_m$ denote the fading coefficients of channel between $m$-th sector of \SUTx ~and PU, and between $m$-th sector of \SUTx ~and \SURx, respectively, when the ESPAR antenna of \SUTx ~is in directional mode, where  $\psi_m = h  \sqrt{p_m(\phi_\text{PU} )} $, $\chi_m= h_{\text{ss}} \sqrt{p_m(\phi_\text{SR} )}$. 
We assume the channel gain $\nu_m = |\chi_m|^2$ is an exponential random variable with mean $\delta_m$, and \SUTx ~knows $\delta_m$, for all $m$ \cite{ArslanESPAR}. For the readers' convenience, we have collected the most commonly used symbols in Table \ref{table1}.
%
%
%
%
%
%
\begin{figure}[!t]
\vspace{-8mm}
\centering
\hspace{0mm}
\setlength{\unitlength}{3.2mm} 
\centering
\scalebox{0.75}{
\begin{picture}(27,9)
\hspace{0mm}
\centering
\put(-2.55,-0.5){\line(0,1){5}}
\put(-2.5,.5){\framebox(32,4){}}
\put(-1.7,2.5){Channel Sensing}
\put(-1.5,1.3){and Monitoring}
\put(6.5,-0.5){\line(0,1){5}}
\put(7.0,2.5){Channel}
\put(7.0,1.3){Training}
\put(11.0,-0.5){\line(0,1){5}}
\put(16,2.1){Data Transmission}
\put(29.52,-0.5){\line(0,1){5}}
\put(-2.55,5.0){\vector(1,0){32.1}}
\put(-2.55,5){\vector(-1,0){0}}
\put(13,5.3){$T_\text{f}$}
\put(-2.55,-0.35){\vector(1,0){9.1}}
\put(-2.55,-0.35){\vector(-1,0){0}}
\put(1.5,-1.5){$T_\text{sen}$}
\put(6.5,-0.35){\vector(1,0){4.5}}
\put(6.5,-0.35){\vector(-1,0){0}}
\put(8,-1.5){$T_\text{train}$}
\put(11.0,-0.35){\vector(1,0){18.5}}
\put(11.0,-0.35){\vector(-1,0){0}}
\put(17,-1.5){${ T_\text{f} - T_\text{sen} - T_\text{train}}$}
\end{picture}}
\vspace{4.5mm} 
\caption{The structure of frame employed by \SUTx.} 
\label{FrameStructure}
\vspace{1mm} 
\end{figure}
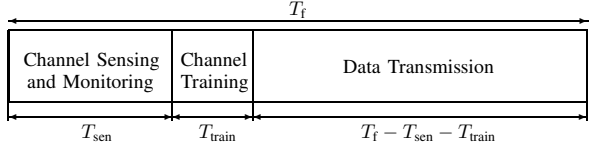
%
%
%
\begin{table}[h!]
  \begin{center}
    \caption{Most commonly used symbols.}
    \vspace{-1mm}
    \label{table1}
    \begin{footnotesize}
    \begin{tabular}{l|l} 
      \textbf{Symbol} & \textbf{Description} \\
      \hline \hline
      $M$ & Number of beams\\
      $N$ & Number of samples used for sensing and monitoring \\
      $n_b$ & Number of bits for quantization at \SURx \\                
      $p_m(\phi)$ &  Radiation pattern of $m$-th beam at angle $\phi$ \\       
      $\psi_m$ &  Fading coefficient of channel between $m$-th beam of \\ & \SUTx ~and PU \\                        
      $\chi_m$ &  Fading coefficient of channel between $m$-th beam of \\ & \SUTx ~and \SURx \\  
      $\delta_m$ &  Mean of channel gain  between $m$-th beam of \\ & \SUTx ~and \SURx \\   
      $\nu^*$ &  Channel gain of selected beam for data transmission \\ & from  \SUTx ~to \SURx \\          
      $\phi_\text{PU}, \phi_\text{SR}$ & Directions of PU and \SURx ~with  respect to \SUTx \\ 
      $m_\text{PU}^*, m_\text{SR}^*$ \!\!\!\!\! & Indices of selected beam for  PU and \SURx \\
      $\pi_0, \pi_1$ & Prior probabilities of $\mathcal{H}_0$  and $\mathcal{H}_1$ \\ 
      $\pioh, \pilh$ &  Probabilities of channel being sensed idle or busy \\       
      $T_\text{f}$ &  Duration of frame employed by \SUTx \\
      $T_\text{sen}$ & Duration of {\it{channel sensing and monitoring phase}} \\
    \end{tabular}
    \end{footnotesize}   
   \end{center}     
   \vspace{-6.5mm}  
\end{table}
%
%
%
\subsection{Our Problem Statement}\label{ProblemStatement}
%
%
Suppose, SUs employ a frame with a fixed duration of $T_\text{f}$ seconds, depicted in Fig. \ref{FrameStructure}. We assume \SUTx ~first senses the channel and monitors the activity of PU. We refer to this period as {\it{channel sensing and monitoring phase}} (with a variable duration of $T_\text{sen}$ seconds). Depending on the outcome of this phase, \SUTx ~stays in this phase or enters the next phase, which we refer to as  {\it{transmission phase}}. The transmission phase itself consists of two phases: {\it{channel training phase}} (with a fixed duration of $T_\text{train}$  seconds) followed by {\it{data transmission phase}} (with a variable duration of $T_\text{f} \!-\! T_\text{sen} \!-\! T_\text{train}$ seconds). During the former phase, \SUTx ~sends pilot symbols to enable channel training and estimation at \SURx. During the latter phase, \SUTx ~sends data symbols to \SURx. Given $T_\text{f}$ and $T_\text{train}$  we have $0< T_\text{sen} < (T_\text{f} -T_\text{train})$. In the following, we describe how \SUTx ~operates in directional mode during these three distinct phases. Based on these descriptions,  we provide our problem statement.
%
%
\par $\bullet$ \textbf{Channel Sensing and Monitoring Phase:} During this phase \SUTx ~senses the channel and monitors the activity of PU. Suppose $\mathcal{H}_1$  and $\mathcal{H}_0$  represent the binary hypotheses of PU being active and inactive, respectively, with prior probabilities $\Pr \{\mathcal{H}_1\} \! = \!  \pi_1$ and $\Pr \{\mathcal{H}_0\} \! =\! \pi_0$. \SUTx ~applies a binary detection rule, as will be described in Section \ref{GLRT_SS}, to decide whether or not PU is active. Let $\widehat{\mathcal{H}}_1$ and $\widehat{\mathcal{H}}_0$ denote the detector outcome, i.e., the detector finds PU active (channel is sensed busy and occupied) and inactive (channel is sensed idle and unoccupied and thus can be used by \SUTx ~for transmission), respectively. The accuracy of this binary detector is characterized  by its false alarm probability $ {{P} }_\text{fa} =\Pr \{\widehat {\mathcal{H}}_1 | \mathcal{H}_0\}$ and detection probability $ {{P} }_\text{d} =\Pr \{\widehat{\mathcal{H}}_1 | \mathcal{H}_1\}$. Therefore, the probabilities of events $\widehat{ \mathcal{H}}_0$ and $\widehat {\mathcal{H}}_1$ become $\pioh \! =\!  \Pr \{ \widehat{ \mathcal{H}}_0 \} \! =\!  \pi_1 (1 \! -\! {P}_\text{d} ) + \pi_0 (1 \! - \! {P}_\text{fa}) $ and $\pilh \! = \! \Pr \{ \widehat{ \mathcal{H}}_1 \} \! = \! \pi_1 {P}_\text{d}  + \pi_0 {P}_\text{fa} $, respectively. Furthermore, the joint probabilities are $\alpha_0\! =\! \Pr \{ \mathcal{H}_0 , \widehat{ \mathcal{H}}_0\} \! =\! \pi_0 (1 \! - \! {P}_\text{fa})$ and $\beta_0 \! = \! \Pr \{\mathcal{H}_1 ,\widehat{ \mathcal{H}}_0\} \! = \! \pi_1 (1 \! -\! {P}_\text{d})$. The accuracy of channel sensing impacts the maximum information rate that \SUTx ~can transmit reliably to \SURx. Our problem formulation incorporates the effect of imperfect channel sensing on the constrained ergodic capacity maximization. As long as the channel is sensed busy, \SUTx ~stays in {\it{channel sensing and monitoring phase}}. While being in this phase, \SUTx ~determines the beam corresponding to the location (orientation) of  PU based on the received signal energy. We denote the sector index corresponding to PU's location by $m_\text{PU}^*$. 
\SUTx ~uses $m_\text{PU}^*$ for adapting its discrete power level during {\it{data transmission phase}}. We note that, there is a non-zero error probability when \SUTx ~determines the beam index $m_\text{PU}^*$, i.e., it is possible that $m_\text{PU}^*$ is not the true beam index corresponding to PU. Our problem formulation takes into account the impact of this error probability on the constrained ergodic capacity maximization. 
%
%
\par $\bullet$ \textbf{Channel Training Phase:} When the channel is sensed idle, \SUTx ~leaves {{\it channel sensing and monitoring phase}} and enters this new phase and sends pilot symbols over all beams. Based on the  received training signal, \SURx ~estimates the channel gain $\nu_m = |\chi_m|^2$ for all beams and determines the strongest channel  $\nu^*=\max \{\nu_m\}$ among all beams, and the corresponding beam index $m_\text{SR}^* = \arg \max \{\nu_m\}$. 
Also, \SURx ~employs an $n_b$-bit quantizer to quantize $\nu^*$. The quantizer has $N_b=2^{n_b}$ thresholds, denoted by $\{\mu_k\}_{k=1}^{N_b}$, satisfying  $\mu_0=0 < \mu_1 < \ldots < \mu_{N_b+1}=\infty$, and has $N_b+1$ quantization intervals $\mathcal{I}_k=[\mu_k, \mu_{k+1})$ for $k=0,\ldots ,N_b$. The quantization mapping rule follows: if the quantizer input $\nu^*$ lies in the interval $\mathcal{I}_k$ then the quantizer output is $\mu_k$, for $k=0, \ldots , N_b$. The index of quantization interval $k$ can be represented by $n_b$-bits. Then, \SURx ~feeds back $m_\text{SR}^*$ as well as the $n_b$-bit  representation of the index of the quantization interval to which $\nu^*$ belongs, over an error-free bandwidth limited feedback link to \SUTx, so \SUTx ~can optimally adapt its discrete power level accordingly. We take into account the probability of determining the true beam corresponding to \SURx ~as well as the probability of selecting the true strongest channel among all beams between \SUTx ~and \SURx, on the constrained capacity maximization. 
%
%
\par $\bullet$ \textbf{Data Transmission Phase:} After {\it{channel training phase}}, \SUTx ~enters {\it{data transmission phase}} and transmits data to \SURx ~over the selected beam $m_\text{SR}^*$. During this phase, \SUTx ~adapts its discrete  power level $P_k$, where $P_k \in \{P_0, P_1, P_2, ..., P_{N_b} \}$, using $m_\text{PU}^*$ and the information received from \SURx ~through the feedback channel, such that the ergodic capacity of \SUTx-\SURx ~link is maximized, subject to 
average interference and transmit power constraints. We let $P_0=0$ to indicate that when $\nu^* \in \mathcal{I}_0=[0,\mu_1)$ then \SUTx ~does not transmit data to \SURx, since the channel is too weak. 
{\color{blue} \par Table \ref{tab:Summery1} enumerates the sequential steps we take within each of the three phases: {\it channel sensing and monitoring phase}, {\it channel training phase}, and {\it data transmission phase}.}
%
%
%
\begin{table*}
	{\color{blue}
    \centering
	\vspace{-1mm}
    \caption{}
	\vspace{-1mm}
    \label{tab:Summery1}
    \begin{footnotesize}
    \begin{tabular}{|l|l|}
        \hline
        \textbf{Phase} &  \textbf{Sequential steps in each phase} \\ 
        \hline
        \multirow{5}{*}{1. Channel Sensing   and Monitoring Phase } & 1.1. \SUTx ~senses the channel and monitors the activity of PU.\\    
       & 1.2. As long as the channel is sensed busy, \SUTx ~stays in this phase.\\
      	& 1.3. While being in this phase, \SUTx ~determines the beam corresponding to the orientation of PU    \\ &  $\quad$ ~ denoted by $m^*_\text{PU}$ (based on the received signal energy). \\
    	   & 1.4. When the channel is sensed idle, \SUTx ~leaves this phase and enters the next phase.\\ 
	\hline
	\multirow{6}{*}{2. Channel Training Phase} & 2.1. \SUTx ~sends pilot symbols over all beams. \\
	& 2.2.  \SURx ~estimates the channel gain $\nu_m$ for all beams and determines the strongest channel $\nu^*$ among \\ &  $\quad$ ~   all beams and the corresponding beam index $m_\text{SR}^*$. \\
	& 2.3. \SURx ~employs an $n_b$-bit quantizer to quantize $\nu^*$.  \\
	& 2.4. \SURx ~feeds back $m_\text{SR}^*$ as well as the $n_b$-bit  representation of the index of the quantization interval  \\ & $\quad$ ~ to which $\nu^*$ belongs, over a feedback link to \SUTx. \\
	& 2.5 \SUTx ~leaves this phase and enters the next phase. \\
	\hline
	\multirow{4}{*}{3. Data Transmission Phase} & 3.1. \SUTx ~adapts its discrete  power level $P_k$, using $m_\text{PU}^*$ and the information received from \SURx ~ \\ & $\quad$ ~  through the feedback channel, such that the constrained ergodic  capacity is maximized. \\
	& 3.2. \SUTx ~transmits data to \SURx ~with power $P_k$ over the selected beam $m_\text{SR}^*$. \\
	\hline
    \end{tabular}
    \end{footnotesize}
    }
    \vspace{-3mm} 
\end{table*}
%
%
{\color{blue} \par {\bf Remark}: It is worth emphasizing that in our problem, \SUTx ~does not know the angles  $\phi_\text{PU}$ and  $\phi_\text{SR} $, defined in Section \ref{Section2B} (otherwise, the beam selection at \SUTx ~for data transmission would become trivial). We take full advantage of beam steering capability of the ESPAR antenna that enables sector-based spectrum  sensing and communication at \SUTx. In this work, \SUTx ~does not estimate the angles $\phi_\text{PU}$ and $\phi_\text{SR}$. Instead it determines the indices of the sectors corresponding to PU and \SURx ~(i.e., \SUTx ~finds $m^*_\text{PU}$ and learns  $m^*_\text{SR}$ during {\it channel sensing and monitoring phase} and {\it channel training phase}, respectively).} For mathematical tractability, we assume that these sectors are unchanged during a frame duration. Comparing with a CR system design that is based on angle (or directional of arrival) estimation at \SUTx, using the sector-based sensing and communication  improves the system design resilience against the mobility of users (as long as the determined sectors do not change due to mobility). 
\par When spectrum sensing is imperfect,  the capacity of \SUTx-\SURx ~link can be written as \cite{CISS2019}
\vspace{-1mm}
\begin{equation}\label{C_Ergodic}
C= D_t \,\mathbb{E} \big \{ \alpha_{0} C_{0,0} + \beta_0 C_{1,0}  \big \}, 
\vspace{-1mm}
\end{equation}
%
%
where $C_{i,0}$ is the instantaneous capacity of this link corresponding to the event $\mathcal{H}_i$ and $\widehat{\mathcal{H}}_0$, $D_t = (T_\text{f}-T_\text{sen}-T_\text{train})/T_\text{f}$ and $\mathbb{E}\{\cdot\}$ is the statistical expectation operator. Let $\Ibar$ indicate the maximum allowed interference power imposed on PU and $\Pbar$ denote the maximum allowed average transmit power of \SUTx. Given our aforementioned system model description and to enable mathematically expressing the average interference and transmit power constraints 
in our problem, we let $P(\nu^*)$ indicate \SUTx ~transmit power in terms of the channel gain of the selected beam $\nu^*$ between \SUTx ~and \SURx. To satisfy the average interference constraint ,
we have 
\vspace{-1mm}
\begin{equation}\label{Iav0}
D_t \beta_0 \,\mathbb{E}  \big\{ g_\text{sp}  ~p( \kappa_{\text{SR} }^*  -  \kappa_{\text{PU}}^*) P(\nu^*) \big \} \leq \Ibar,
\vspace{-1mm}
\end{equation}
%
and to satisfy  the average transmit power constraint, 
we have
%
\begin{equation}\label{Pav0}
 D_t \pioh \,\mathbb{E} \big\{  P(\nu^*) \big\} \leq \Pbar.  
\end{equation}
%
Notice that, had channel sensing have been ideal, $\beta_0=0$ and data communication between \SUTx ~and \SURx ~would cause no interference on PU. The more accurate channel sensing is, the smaller is the power of interference signal imposed on PU. On the other hand, increasing the accuracy of channel sensing requires a longer $T_\text{sen}$ and a shorter $D_t$, given the frame duration $T_\text{f}$. Reducing $D_t$ decreases the capacity $C$ in \eqref{C_Ergodic}. Therefore, there is a tradeoff between increasing $C$ and decreasing the power of interference signal imposed on PU. Let $F_{\nu^*}(\cdot)$ be the cumulative distribution function (CDF) of $\nu^*$ (will be derived in Section \ref{SubSe3}). Given the discrete power levels $P_k$'s and the quantization thresholds $\mu_k$'s, $\mathbb{E} \left\{  P(\nu^*) \right\} $ can be written as
\vspace{-2mm}
\begin{equation}
 \mathbb{E} \big\{ P(\nu^*) \big \} = \sum_{k=1}^{N_b}  P_k \Big [ F_{\nu^* }(\mu_{k+1})  -  F_{\nu^* }(\mu_{k})  \Big ].
\vspace{-1mm}
\end{equation}
%
Therefore, the constraints in \eqref{Iav0} and \eqref{Pav0} can be rewritten as 
\vspace{-2mm}
%
\begin{equation}\label{Iav01}
D_t \beta_0 \gamma_\text{sp} \,\mathbb{E} \big\{ p( \kappa_{\text{SR}}^*  \! -  \! \kappa_{\text{PU}}^* ) \big \}  \sum_{k=1}^{N_b}\!   P_k \Big [ F_{\nu^* }\! (\mu_{k+1})  \! -  \! F_{\nu^* }\! (\mu_{k})  \Big ]  \leq \Ibar, 
\vspace{-2mm} 
\end{equation}
%
%
\vspace{-4mm}
\begin{equation}\label{Pav01}
D_t \pioh  \sum_{k=1}^{N_b}  P_k \Big [ F_{\nu^* }(\mu_{k+1}) -  F_{\nu^* }(\mu_{k})  \Big ]   \leq \Pbar.  
\vspace{-1mm}
\end{equation}
%
\noindent Our main objective is to find the optimal channel sensing and monitoring duration $T_\text{sen}$, the optimal quantization thresholds $\mu_k$'s for the channel gain quantizer employed at \SURx, and the optimal discrete power levels $P_k$'s corresponding to each quantization interval  $I_k=[\mu_k, \mu_{k+1})$, such that the ergodic capacity $C$ in \eqref{C_Ergodic} is maximized, subject to average interference and transmit power constraints 
given in \eqref{Iav01} and \eqref{Pav01}, respectively. In other words, we are interested in solving the following constrained optimization problem
%
%
\leqnomode
\begin{align*}\tag{P1}\label{Prob1}
{\underset { T_\text{sen}, \{\mu_k , P_k \} _{k=1}^{N_b}}  {\text{Maximize}} }  ~C = D_t \,\mathbb{E} \big \{ \alpha_{0} C_{0,0} + \beta_0 C_{1,0}  \big \}& \nonumber \\
\end{align*}
\vspace{-12mm}
\begin{align}
\begin{array}{ll}
\text{s.t.:}  ~~& 0  < T_\text{sen} < (T_\text{f}  \!-\!T_\text{train}),\nonumber \\
 &  0 < \mu_1< \ldots < \mu_{N_b} < \infty,  \nonumber \\
 &   ~P_k > 0 ~\forall k, \nonumber \\
 & \eqref{Iav01}  ~\text{and} ~ \eqref{Pav01} ~\text{are satisfied.} \nonumber
 \end{array}
 \vspace{0mm}
\end{align}
\reqnomode
%
%
%
%
\section{Characterizing Objective Function and Constraints in \eqref{Prob1} }\label{Sect3}
\par Characterizing the objective function and the constraints in \eqref{Prob1} requires addressing the following three components. First, the performance of the binary detector employed by \SUTx ~to detect PU activity during {\it{channel sensing and monitoring phase}} plays role in the objective function and the  average interference constraint 
 in \eqref{Iav01} via $\beta_0$, and in the  average transmit power constraint 
 in \eqref{Pav01} via $\pioh$. Obviously, this performance depends on the choice of the detector. Section \ref{GLRT_SS} describes our proposed binary detector, which is based on the energy of the collected measurements from all sectors of the ESPAR antenna at \SUTx ~during this phase, and provides closed form expressions for ${P}_\text{d}$ and ${P}_\text{fa}$ of this detector. Second, the error probability of finding the sector index $m^*_\text{PU}$ corresponding to PU at \SUTx ~during {\it{channel sensing and monitoring phase}} affects the  average interference constraint  
 in \eqref{Iav01}. This error probability depends on the mechanism through which \SUTx ~determines this sector index. Section \ref{PU_Beam_Selection} explains how \SUTx ~finds this beam index, using the received signal energy from all sectors of the ESPAR antenna during this phase, and derives closed form expression of the corresponding error probability. Third, the  probability of finding the sector index $m^*_\text{SR}$ corresponding to \SURx ~during {\it{channel training phase}} impacts the average interference constraint 
in \eqref{Iav01}. During {\it{data transmission phase}} \SUTx ~sends data to \SURx ~over the selected beam $m^*_\text{SR}$. Section \ref{SubSe3} discusses the method utilized by \SURx ~to find this beam index, using the  received training signal transmitted by all sectors of \SUTx ~antenna, and derives a closed form expression for the corresponding probability. 
%
%
\subsection {Energy-Based Binary Detector for Channel Sensing Using ESPAR Antenna}\label{GLRT_SS}
%
%
\par Channel sensing at \SUTx ~(detecting the activity of PU) during {\it{channel sensing and monitoring phase}}  can be formulated as a binary hypothesis testing problem. Suppose when PU is active (present), it transmits signal $s(t)$ with power $P_\text{p}$. Let $y_m(n)$ denote the discrete-time representation of received signal at $m$-th sector of \SUTx ~at time instant $t=nT_\text{s}$ where $T_\text{s}$ is the sampling period.  Assuming \SUTx ~collects $N= \lfloor  T_\text{sen}/(M T_\text{s}) \rfloor$ samples corresponding to each sector we can write
\vspace{-2mm}
\begin{align}
y_m(n) = & \psi_m (n) s(n) + w_m(n),  \\
\text{for} ~~n=1   \!+\!(m & \!-\!1)N,  \ldots, mN  ~~~~~~~m=1, \ldots, M \nonumber
\end{align}
%
We model the  transmitted signal $s(n)$ by PU as a zero-mean complex Gaussian random variable with variance $P_\text{p}$ and we assume \SUTx ~knows $P_\text{p}$. The term $w_m(n)$ is the additive noise at $m$-th sector of \SUTx ~antenna and is modeled as 
$w_m(n) \sim \mathcal{CN}(0,\sigma_\text{w}^2) $. We assume that $\psi_m(n)$, $s(n)$ and $w_m(n)$ are mutually independent random variables. Since \SUTx ~takes samples of the received signal for different sectors sequentially (in different time instants), $\psi_m(n)$ and noise samples $w_m(n)$ are independent and thus uncorrelated  both in time and space (sector) domains. Under hypothesis $\mathcal{H}_1$, given $\psi_m$, we have $y_m(n) \sim \mathcal{CN}( 0, \sigma_m^2 \!+\! \sigma_\text{w}^2)$ where $\sigma^2_{m} = |\psi_m|^2 P_\text{p}$. Under  hypothesis $\mathcal{H}_0$, we  have $y_m(n)  \sim \mathcal{CN} (0,\sigma_\text{w}^2)$. The hypothesis testing problem at discrete time instant $n$ for $m$-th sector is then given by
\vspace{-1mm}
\begin{equation} 
\begin{cases}
 {\cal H}_{0}: &  y_m(n) =  w_m(n), \\
 {\cal H}_{1}: & y_m(n) =  \psi_m(n)  s(n) + w_m(n). 
\end{cases}
 \vspace{-1mm}
\end{equation}
%
%
Our proposed energy-based binary detector uses all the collected samples from $M$ sectors (total of $N_\text{eq}=MN$ collected samples). Let $\varepsilon_m$ be the energy of received signal at sector $m$. We have
\vspace{-1mm}
\begin{equation}
\varepsilon_m = \frac{1}{N} \! \sum_{n=1\!+\!(m\!-\!1)N}^{mN} \!\!\!  \big | y_m(n) \big |^2.  
\vspace{-1mm}
\end{equation}
%
Under hypothesis $\mathcal{H}_0$ and also under $\mathcal{H}_1$ (given $\psi_m$), the sector energy $\varepsilon_m$ is distributed as a central chi-square random variable with $2N$ degrees of freedom. We consider the summation of energies of received signals over all sectors as the decision statistics $T$ given below
\vspace{-2mm}
\begin{equation}
T = \frac{1}{M} \sum_{m=1}^M \varepsilon_m 
\gtreqless
\begin{matrix}
\widehat{\mathcal{H}}_1 \cr \widehat{\mathcal{H}}_0
\end{matrix}
\eta. 
\vspace{-1mm}
\end{equation}
%
where $\eta$ is the decision threshold. We can rewrite $T$ as
\vspace{-1mm}
\begin{equation}
T = \frac{1}{M N} \sum_{m=1}^{M} \sum_{n=1\!+\!(m\!-\!1)N}^{mN} \!\!\!  \big | y_m(n) \big |^2. 
\vspace{-1mm}
\end{equation}
%
Note that $T$ is the summation of $N_\text{eq}$ random variables. When $N_\text{eq}$ is large enough $T$ can be approximated  as a Gaussian random variable. Thus, Under hypothesis $\mathcal{H}_0$, for large $N_\text{eq}$ we invoke the central limit theorem (CLT), to approximate $T$ as Gaussian with distribution $T \sim  \mathcal{N}  ( \sigma_\text{w}^2, \sigma^2_{T|{\cal H}_0} )$, where $\sigma^2_{T|{\cal H}_0} = {\sigma_\text{w}^4}/N_\text{eq}$. Similarly, under hypothesis $\mathcal{H}_1$ for large $N_\text{eq}$, $T$ can be approximated with another Gaussian with distribution $T \sim  \mathcal{N}  (\zeta, \sigma^2_{T|{\cal H}_1})$ where $\zeta = P_\text{p} \gamma E_A + \sigma_\text{w}^2$, and $\sigma^2_{T|{\cal H}_1}$ is given below
\vspace{-2mm}
\begin{align}\label{sigma2_H1}
\sigma^2_{T|{\cal H}_1} = & \frac{1}{N_\text{eq}} \Big [ \sigma_\text{w}^4 + 2 \gamma P_\text{p} E_A \sigma_\text{w}^2 +\gamma^2 P_\text{p}^2 \big (3 E_B  -  MN E_A^2 \big ) \Big ] \nonumber \\
 & +  \frac{\gamma^2 P_\text{p}^2}{M^2} \sum_{m=1}^{M} \sum_{m^\prime=1}^{M} E_{m m^\prime},
\end{align}
%
where
\vspace{-3mm}
\begin{subequations}\label{EmmA}
\begin{align}
E_{m m^\prime} = & \,\frac{1}{2 \pi} \int_{0}^{2\pi} \! p_m(\theta) p_{m^\prime}(\theta) d\theta, \label{EmmAa} \\
E_A = & \,\frac{1}{2\pi} \int_{0}^{2\pi} p(\theta)  d\theta, \label{EmmAb}
\end{align}
\end{subequations}
\noindent and $E_B = E_{mm}$. Then, the false alarm and detection probabilities of this detector are given as the following
\vspace{-1mm}
\begin{equation}
\begin{array}{ll}
P_\text{fa} = Q \left(\frac{\eta - \sigma_\text{w}^2}{\sigma_{T|{\cal H}_{0}}} \right ),  ~~~~~~~~& P_\text{d} = Q \left( \frac{\eta -\zeta}{ \sigma_{T|{\cal H}_{1}}} \right ),
\end{array}%
\vspace{-1mm}
\end{equation}
%
where $Q(\cdot)$ is the Q-function. For a given value of $P_\text{d} = \overline{P}_\text{d}$, the false alarm probability can be written as

%
\begin{equation}\label{}
P_\text{fa} = Q \left( \frac{  \sigma_{T|{\cal H}_1} Q^{-1} \big (\overline{P}_\text{d} \big ) + \zeta -\sigma^2_\text{w} }{\sigma_{T|{\cal H}_0 }} \right ).
\end{equation}
%
%
%
%
\vspace{-3mm}
\subsection{Determining the Beam Corresponding to PU}\label{PU_Beam_Selection}
\begin{figure}[!t]
\vspace{-0mm}
\centering
	\begin{subfigure}[b]{0.25\textwidth}                
		\centering	              
		\psfrag{g}[Bl][Bl][0.5]{$g$}
		\psfrag{gss}[Bl][Bl][0.5]{$g_\text{ss}$}
		\psfrag{gsp}[Bl][Bl][0.5]{$g_\text{sp}$}
		\psfrag{SUtx}[Bl][Bl][0.6]{\SUTx}
		\psfrag{SUrx}[Bl][Bl][0.6]{\SURx}
		\psfrag{PU}[Bl][Bl][0.6]{PU}
		\psfrag{phiSR}[Bl][Bl][0.5]{$\phi_\text{SR}$}
		\psfrag{phiPU}[Bl][Bl][0.5]{$\phi_\text{PU}$}
		\includegraphics[width=38mm]{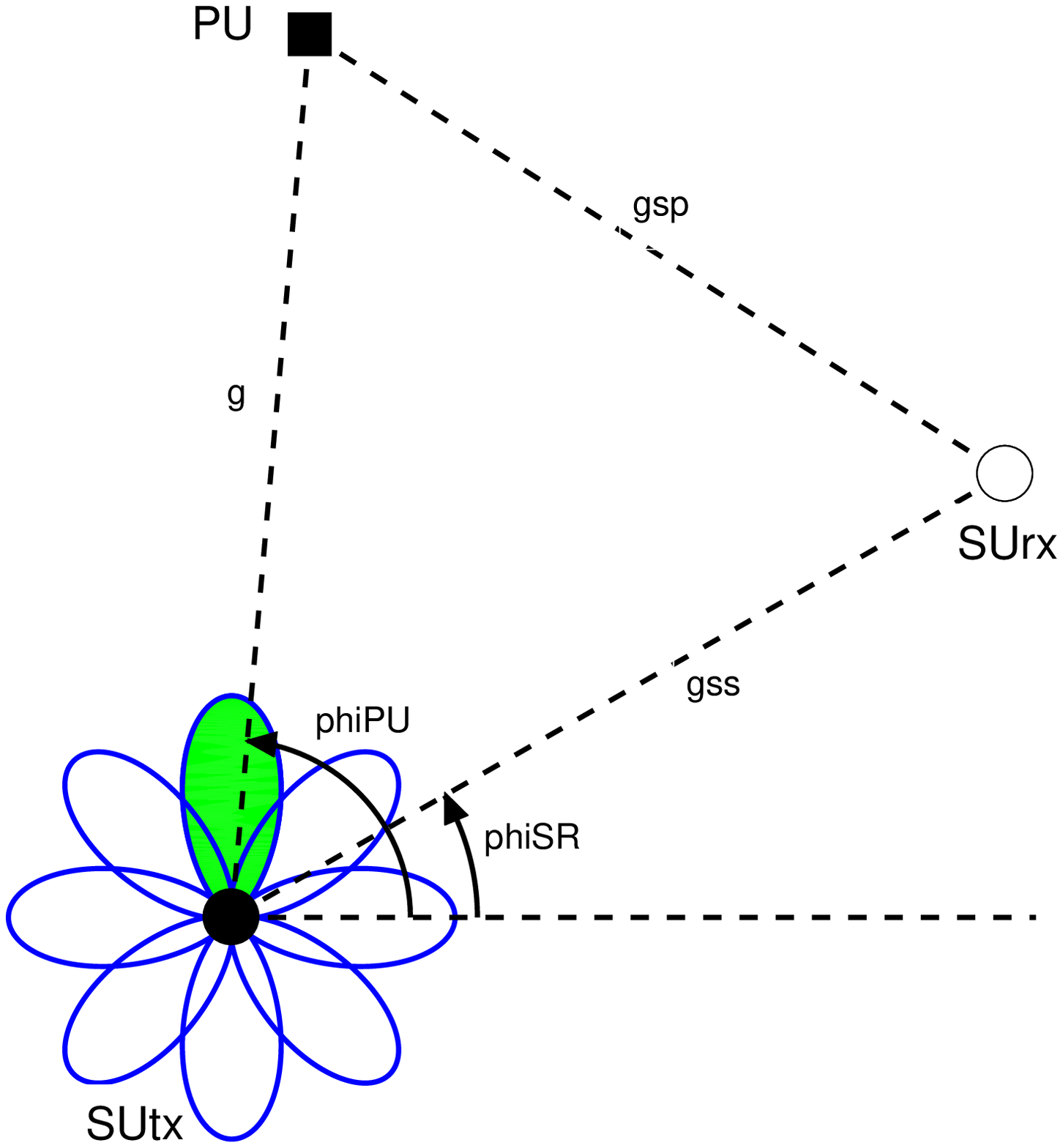}
		\vspace{-4mm}
		\caption{} 
		\label{i_PU}      	       
	\end{subfigure}%
      \begin{subfigure}[b]{0.25\textwidth}
 		\centering      
		\psfrag{g}[Bl][Bl][0.5]{$g$}
		\psfrag{gss}[Bl][Bl][0.5]{$g_\text{ss}$}
		\psfrag{gsp}[Bl][Bl][0.5]{$g_\text{sp}$}
		\psfrag{SUtx}[Bl][Bl][0.6]{\SUTx}
		\psfrag{SUrx}[Bl][Bl][0.6]{\SURx}
		\psfrag{PU}[Bl][Bl][0.6]{PU}
		\psfrag{phiSR}[Bl][Bl][0.5]{$\phi_\text{SR}$}
		\psfrag{phiPU}[Bl][Bl][0.5]{$\phi_\text{PU}$}
		\includegraphics[width=38mm]{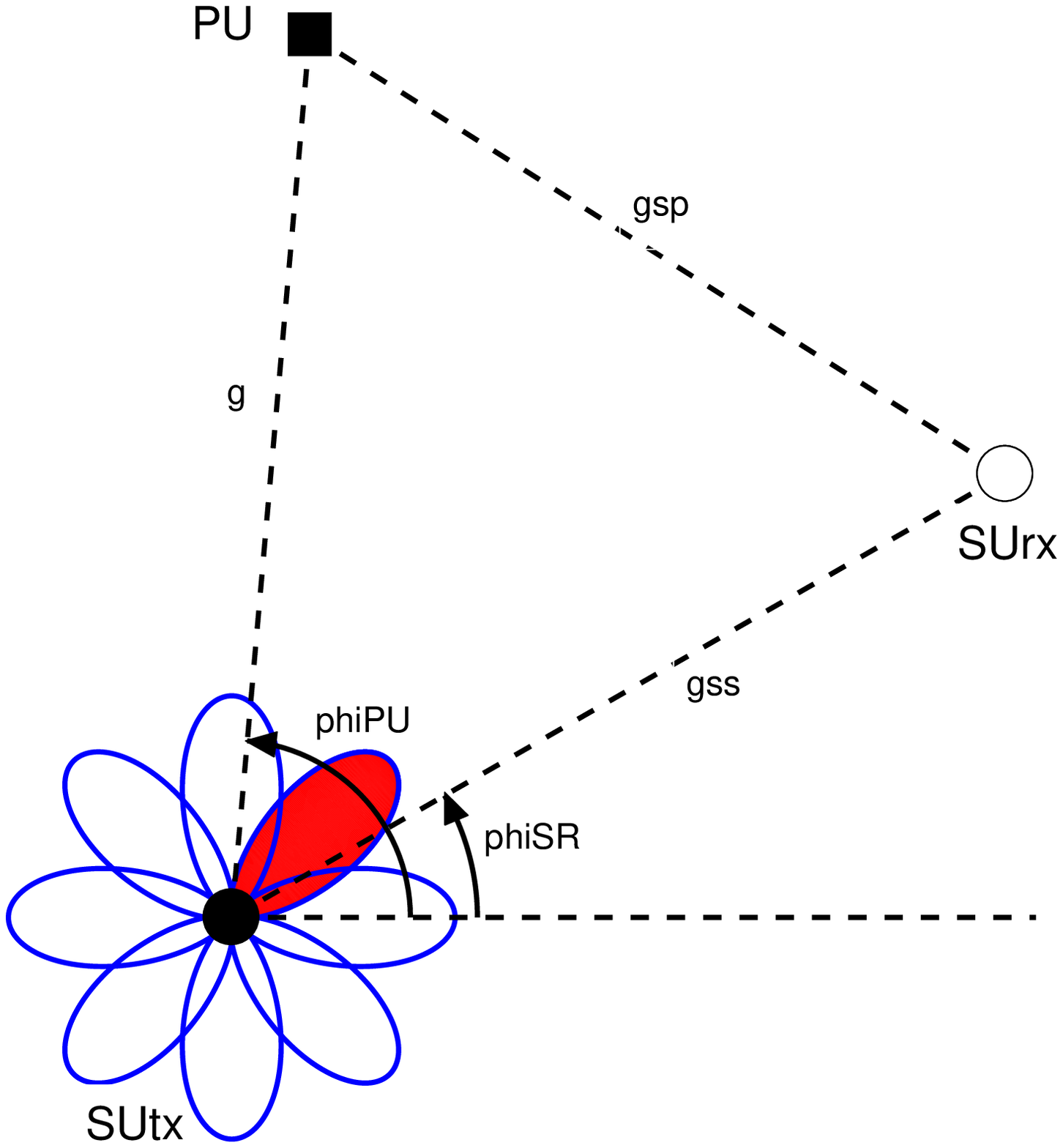}
		\vspace{-4mm}
		\caption{} 
		\label{i_SR}   
      \end{subfigure} \\
      \vspace{-1mm}
\caption{A schematic to show how different beams can be selected to indicate the orientation of \SUTx ~with respect to PU and \SURx ~(a) $m_\text{PU}^*\!=\! 3$, (b) $m_\text{SR}^*\!=\!2$.}
\vspace{-4mm}
\end{figure}

\vspace{-0mm}
During {\it{channel sensing and monitoring phase}} when the channel is sensed busy, \SUTx ~determines the beam corresponding to the orientation of PU based on the received signal energy $\varepsilon_m, m=1, \ldots, M$. Ordering these calculated energies, \SUTx ~selects the beam index corresponding to the largest energy $m_\text{PU}^* = \arg\max \{ \varepsilon_m \}$ among all sectors. For example, in Fig. \ref{i_PU}, we have $m_\text{PU}^*=3$, that is, the third beam has received the largest amount of energy.
%
%
%
%
%
\noindent As we mentioned, under hypothesis $\mathcal{H}_1$, given $\psi_m$ (or equivalently given $g$ and $\phi_\text{PU}$), the sector energy $\varepsilon_m$ is distributed as a central  chi-square random variable with $2N$ degrees of freedom and its conditional pdf and CDF expressions are
%
%
%
\vspace{-1mm}
\begin{subequations}\label{fFeps}
\begin{align}
f_{\varepsilon_m} \big (x  | g, \phi_\text{PU} \big ) = \frac{ x ^{N-1}  ~e^{\frac{-x}{ \sigma_{\text{e}_{m}}^2 } }}{ \sigma_{\text{e}_{m}}^{2N} \Gamma(N)},\\
 F_{\varepsilon_m}\big (x |g, \phi_\text{PU} \big) = \frac{\gamma(N,\frac{x}{\sigma_{\text{e}_m}^2  })}{\Gamma(N) },
\end{align}
\end{subequations}
%
%
\noindent where $\sigma^2_{\text{e}_m} =  (\sigma_m^2 \!+\! \sigma_\text{w}^2 )/{N}$ and $\gamma(\cdot , \cdot)$ is the lower incomplete gamma function
\vspace{-1mm}
\begin{equation}
\gamma(s,x) =  x^{s}  e^{-x} \Gamma(s) \sum_{j=0}^{\infty} \frac{x^j}{\Gamma(j+s+1) }.
\vspace{-1mm}
\end{equation}
%
Let $\overline{\Delta}_{i,m}$ represent the average error probability of finding the sector index corresponding to PU, i.e., the probability that $m_\text{PU}^* =i$ while the true orientation of PU belongs to the angular domain of $m$-th sector, $\phi_\text{PU} \in \Phi_m = \big [\frac{2\pi(m-3/2)}{M} ,\frac{2\pi(m-1/2)}{M}  \big )$, for $i \neq m, i, m =1, \ldots ,M$.  To find $\overline{\Delta}_{i,m}$ we start with finding $\Omega_i= \Pr\{m_\text{PU}^* \!=\! i | g, \phi_\text{PU} \}$, which is the probability that the index of selected sector, given $g$ and $\phi_\text{PU}$, is $i$. We have
\vspace{-1mm}
\begin{align}\label{ProbSelection}
\Omega_i =   &\Pr \big \{m_\text{PU}^*  =  i  \big | g, \phi_\text{PU}  \big \}  \nonumber \\
 =  &\Pr \Big \{ \varepsilon_1<\varepsilon_{i},  \ldots, \varepsilon_{i-1}<\varepsilon_{i}, \varepsilon_{i+1} < \varepsilon_{i}, \ldots, \varepsilon_M<\varepsilon_{i} \Big \} \nonumber   \\
=  &\mathbb{E}_{\varepsilon_i} \left \{ \prod_{\underset{m \neq i} {m=1} }^{M} F_{\varepsilon_m} \big (x | g, \phi_\text{PU} \big)  \right \}  \nonumber   \\
 = & \int_{0}^{\infty}  f_{\varepsilon_i} \big (y  | g,\phi_\text{PU} \big) \! \prod_{\underset{m \neq i} {m=1} }^{M} \! F_{\varepsilon_m} \big (y  | g, \phi_\text {PU} \big) dy.
\end{align}
%
%
\noindent in which $f_{\varepsilon_m} (x  | g, \phi_\text{PU} )$ and $F_{\varepsilon_m} (x  | g, \phi_\text{PU} )$ are the conditional pdf and CDF of $\varepsilon_m$ given in \eqref{fFeps}. Without loss of generality, suppose $i = 1$.  After some mathematical manipulations and taking expectation with respect to $\varepsilon_1$, $\Omega_1$  in \eqref{ProbSelection} can be written as
\vspace{-2mm}
\begin{equation}
\Omega_1  =  \frac{G^{-MN}}{\Gamma(N) \prod_{ {m=1}}^{M} \!\sigma_{\text{e}_m}^{2N} }  ~\widetilde{ \sum\nolimits}_{k_2:k_M }   \!\!\!\frac{\Gamma \big (MN \! + \! \sum_{{j=2}}^{M} k_j \big ) }{E_k ~G^{ \sum_{{j=2}}^{M} k_j } },
\vspace{-2mm}
\end{equation}
%
\noindent where 
\vspace{-1mm}
%
\begin{equation*}
\widetilde{ \sum\nolimits}_{k_2:k_M} =  \sum_{k_2=0}^{\infty}~ \sum_{k_3=0}^{\infty} ... \sum_{k_M =0}^{\infty},
\vspace{-0mm}
\end{equation*}
%
%
\vspace{-2mm}
\begin{equation*}
E_k =  \prod_{{j=2}}^{M}  \sigma_{\text{e}_j}^{2 k_j} \Gamma(k_j+N+1),  ~~~~~~G =  \sum_{m=1}^{M} \frac{1}{\sigma_{\text{e}_m}^2}.
\vspace{-1mm}
\end{equation*}
%
\par
\begin{figure}[!t]
\vspace{-0mm}
\centering
	\begin{subfigure}[b]{0.25\textwidth}                
		\centering	              
		\psfrag{phiPT}[Bl][Bl][0.7]{$\phi_\text{PU}$ [degree]}
		\psfrag{Delta}[Bl][Bl][0.7]{$\Delta_1$}
		\psfrag{phi  =  20}[Bl][Bl][0.36]{$\phi_\text{3dB} \!=\!20\degree$}
		\psfrag{phi  =  30}[Bl][Bl][0.36]{$\phi_\text{3dB}\!=\!30\degree$}
		\includegraphics[width=43mm]{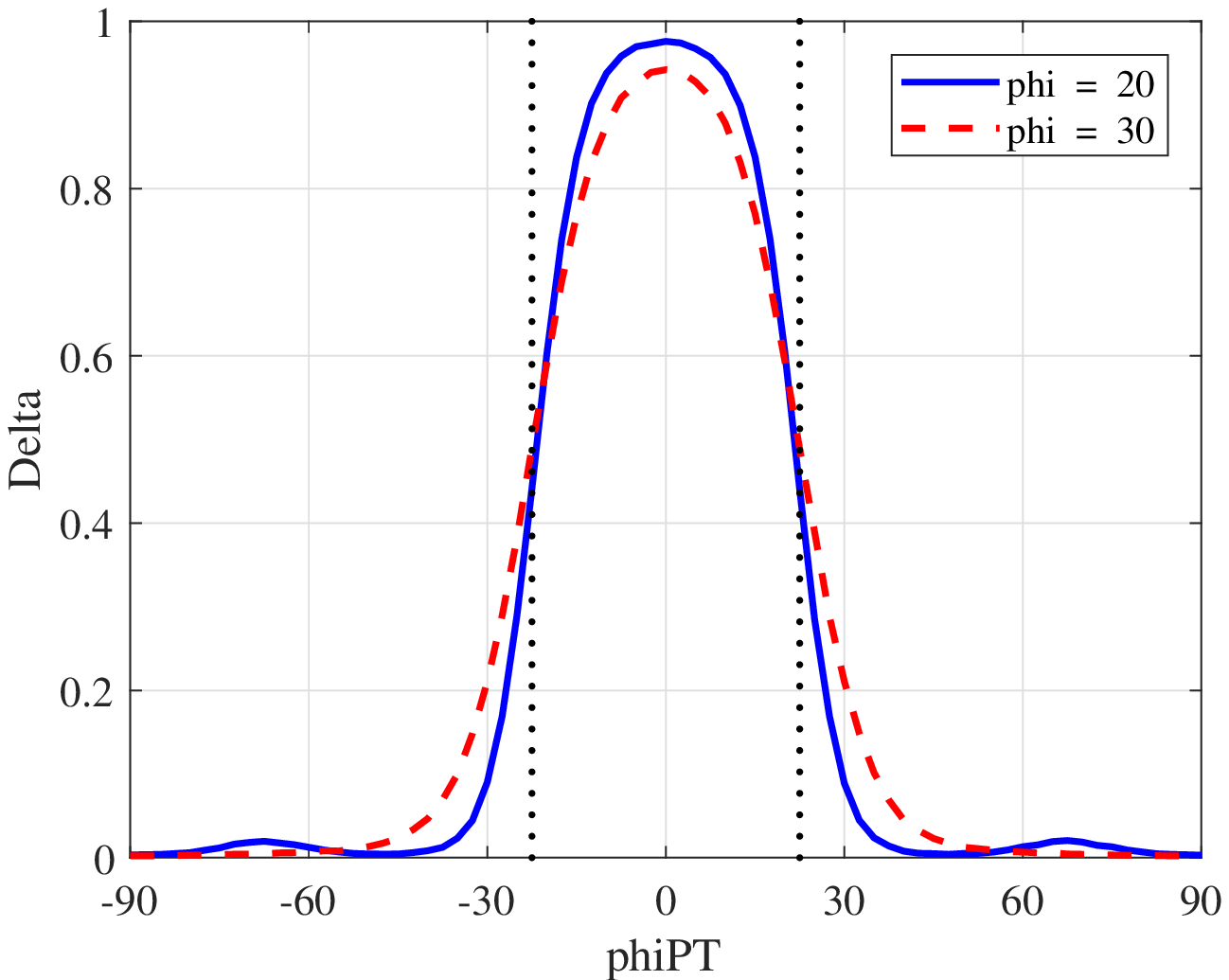}
		\caption{} 
		\label{PU1stSelection2}     	             
	\end{subfigure}%
      \begin{subfigure}[b]{0.25\textwidth}
 		\centering      
		\psfrag{phiPT}[Bl][Bl][0.7]{$\phi_\text{PU}$ [degree]}
		\psfrag{Delta}[Bl][Bl][0.7]{$\Delta_1$}
		\psfrag{N=10}[Bl][Bl][0.34]{$N \!= \! 10$}
		\psfrag{N=30}[Bl][Bl][0.34]{$N \!= \!30$}
		\psfrag{N=200}[Bl][Bl][0.34]{$N \!= \!200$}
		\includegraphics[width=43mm]{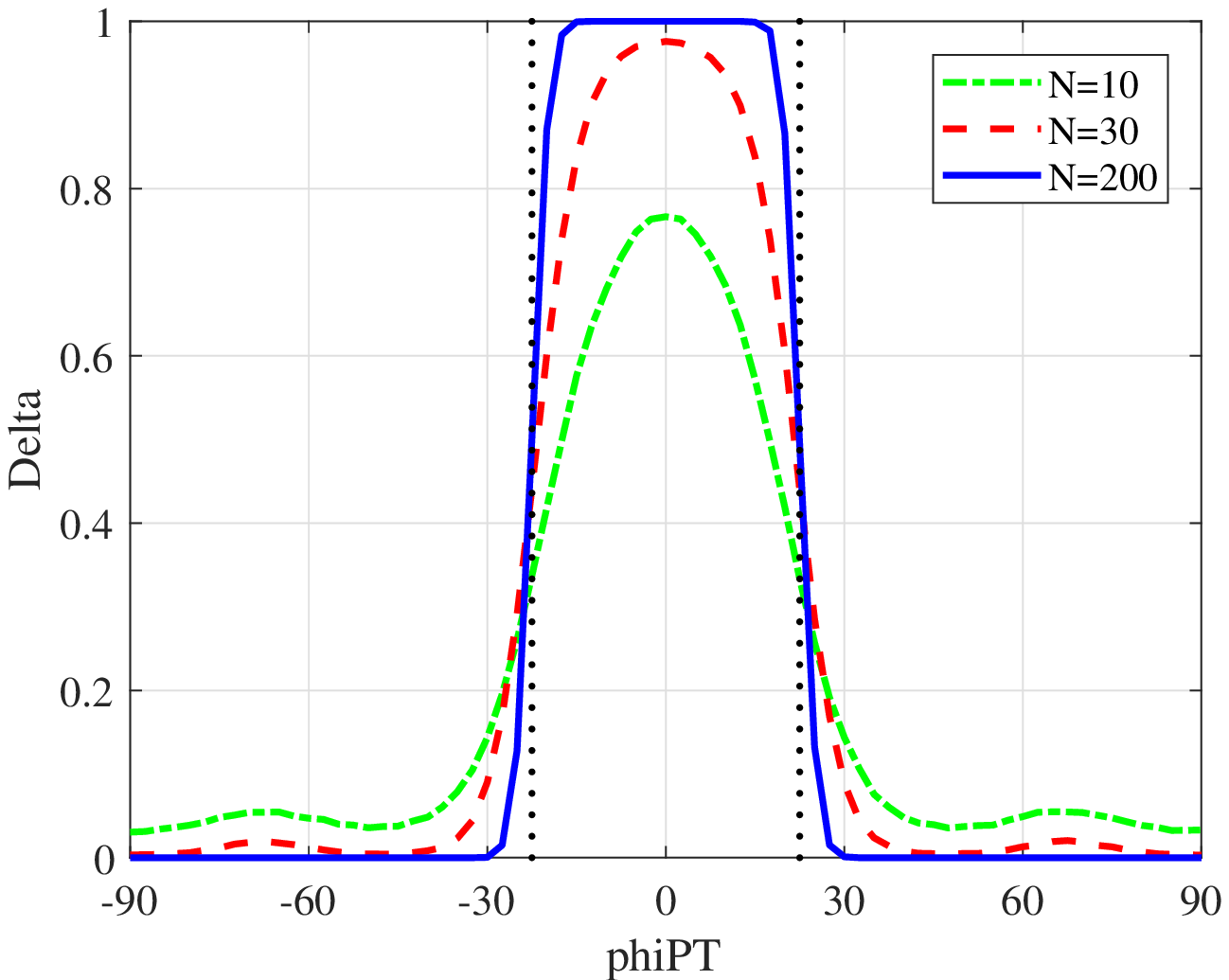}
		\caption{} 
		\label{PU1stSelection5}   
      \end{subfigure} \\
      \vspace{-1mm}
\caption{$\Delta_1$ versus $\phi_\text{PU}$ for $M\!=\!8$ and $\text{SNR}_{\text{PU}}\!=\!0$\,dB (a) $N\!=\!20$, $\phi_{ \text{3dB} }\!=\!20\degree, 30\degree$ (b) $\phi_{ \text{3dB} }\!=\!20\degree$, $N=10, 30, 200$.}
\vspace{-3mm}
\end{figure}

\begin{figure}[!t]
\vspace{-0mm}
\centering
	\begin{subfigure}[b]{0.25\textwidth}                
		\centering	              
		\psfrag{N = 20}[Bl][Bl][0.42]{$N\!=\!20$}
		\psfrag{N = 60}[Bl][Bl][0.42]{$N\!=\!60$}
		\psfrag{N = 100}[Bl][Bl][0.42]{$N\!=\!100$}
		\psfrag{Delta}[Bl][Bl][0.7]{$\overline{\Delta}_{1,m}$}
		\psfrag{pattern index}[Bl][Bl][0.6]{index beam $m$}
		\includegraphics[width=43mm]{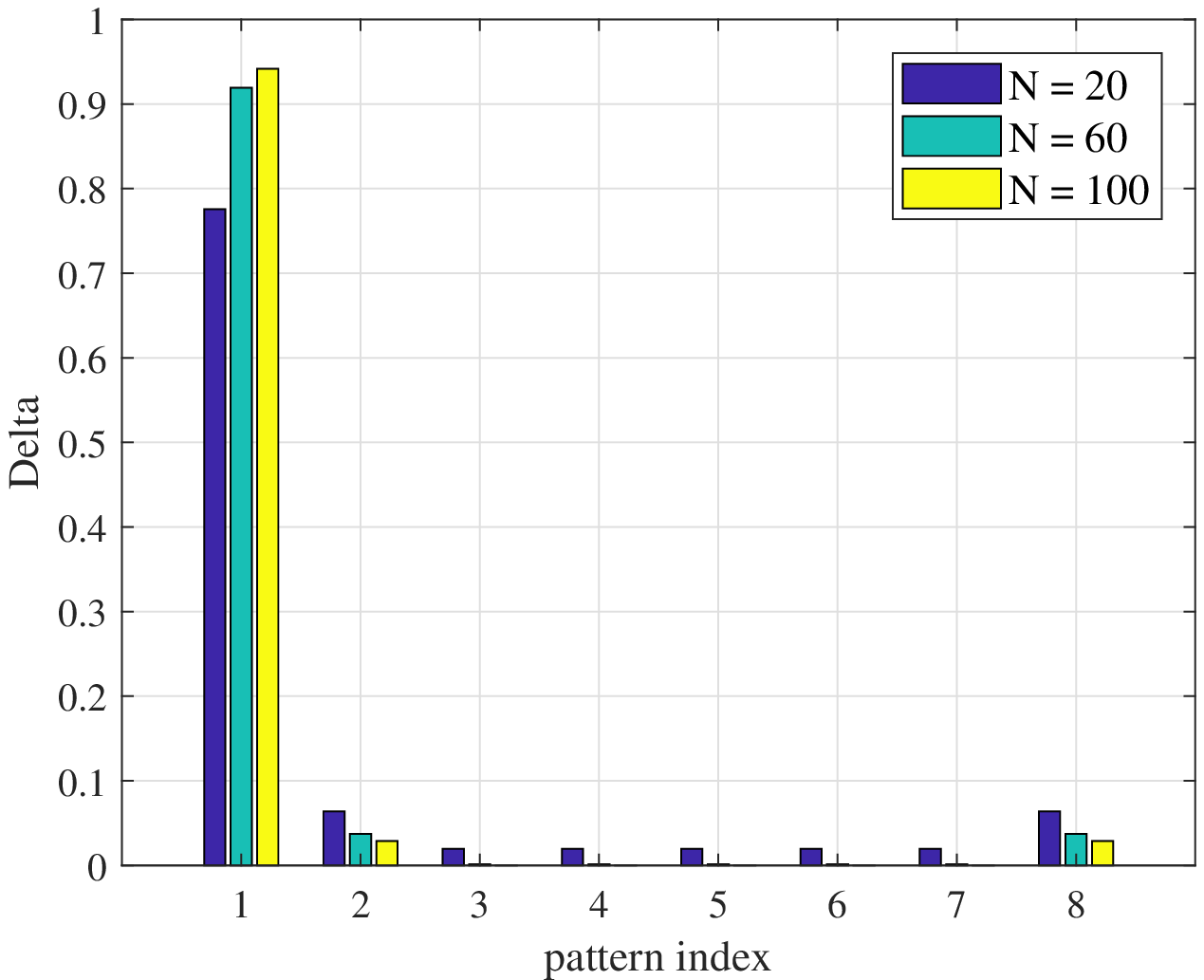}
		\caption{} 
		\label{PUSelection_Bar1}                   
	\end{subfigure}%
      \begin{subfigure}[b]{0.25\textwidth}
 		\centering      
		\psfrag{N = 20}[Bl][Bl][0.42]{$N\!=\!20$}
		\psfrag{N = 60}[Bl][Bl][0.42]{$N\!=\!60$}
		\psfrag{N = 100}[Bl][Bl][0.42]{$N\!=\!100$}
		\psfrag{Delta}[Bl][Bl][0.7]{$\overline{\Delta}_{1,m}$}
		\psfrag{pattern index}[Bl][Bl][0.6]{index beam $m$}
		\includegraphics[width=43mm]{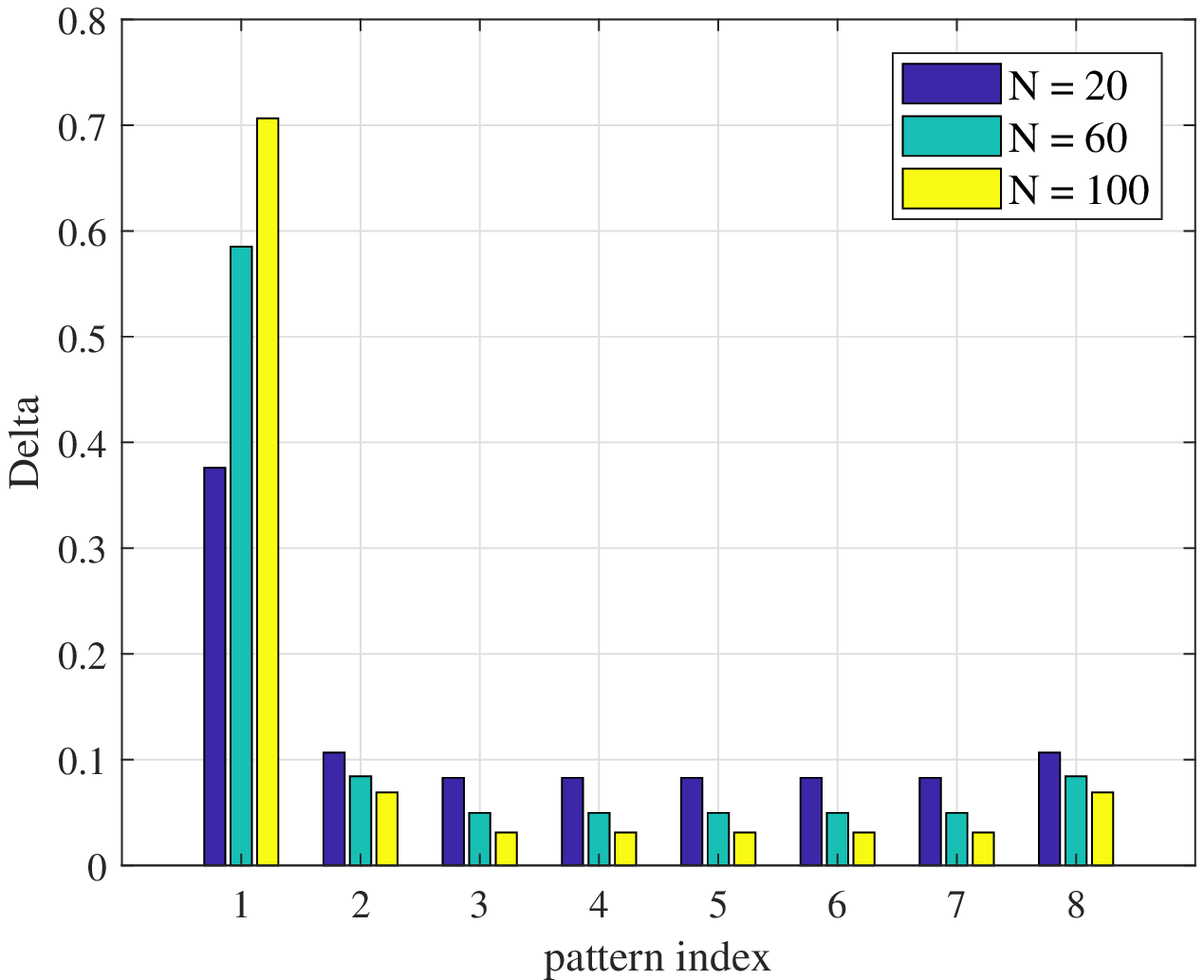}
		\caption{} 
		\label{PUSelection_Bar2}    
      \end{subfigure} \\
      \vspace{-1mm}
\caption{$\overline{\Delta}_{1,m}$ versus the index beam $m$ for $\phi_{3\text{dB}}\!=\!20^\degree$ (a) $\text{SNR}_{\text{PU}} \!= \!0$\,dB, (b) $\text{SNR}_{\text{PU}} \!= \! -5$\,dB.}
\vspace{-3mm}
\end{figure}

 To illustrate the behavior of $\Omega_1$ (averaged over fading gain $g$) we define $\Delta_1=\mathbb{E}_g \{\Omega_1 \} = \Pr\{m^*_\text{PU}\!=\!1 | \phi_\text{PU} \}$ and plot $\Delta_1$ versus $\phi_\text{PU}$ for $M\!=\!8$ and $\text{SNR}_{\text{PU}} \!=\! {\gamma P_\text{p}}/{\sigma_\text{w}^2} \!=\! 0$\,dB. Fig. \ref{PU1stSelection2} shows $\Delta_1$ versus $\phi_\text{PU}$ for $N\!=\!20$ and $\phi_{ \text{3dB} }\!=\!20^\degree, 30^\degree$. We observe  that when $\phi_{ \text{3dB} } $ decreases from $30\degree$ to $20\degree$, beam selection becomes more accurate, i.e.,  $\Delta_1$ increases for $ \phi_\text{PU} \! \in \! \Phi_1 \!=\! [-22.5\degree, 22.5\degree)$, however, it decreases outside this angular interval. Fig. \ref{PU1stSelection5} plots $\Delta_1$ versus $\phi_\text{PU}$ for $N\!=\!10, 30, 200$ and $\phi_{ \text{3dB} } \!=\! 20^\degree$. We observe that as $N$ increases beam selection becomes  more accurate.  For large $N$, we can see that $\Delta_1$ approaches one for $ \phi_\text{PU} \in \Phi_1$ and it is approximately zero outside this angular interval. Now, we are ready to find $\overline{\Delta}_{i,m}$ using $\Delta_i=\Pr \{m^*_\text{PU}=i |\phi_\text{PU} \}$. We have
%
%
%
%
%
\begin{equation}\label{DeltaAverage}
\overline{\Delta}_{i,m} =  \int \nolimits_{ \phi_\text{PU} \in \Phi_m }  \!\!\!\!\! \Delta_i \Pr \! \big \{ \phi_\text{PU} \! \in \! \Phi_m \big \} ~d\phi_\text{PU}. 
\end{equation}
%
Due to the symmetrical structure of the ESPAR antenna we have $\overline{\Delta}_{i,m}=\overline{\Delta}_{m,i}$. Note that $\overline{\Delta}_{i,i}$ is the probability of selecting the correct beam and $\overline{\Delta}_{i,m}$ for $i \neq m$ is the probability of selecting the incorrect beam, leading to error probability in beam selection. The average error probability $\overline{ \Delta}_{1,m}$ versus the index beam $m$ is shown in Figs. \ref{PUSelection_Bar1} and \ref{PUSelection_Bar2} for $\text{SNR}_{\text{PU}}\!=\!0, -5$\,dB. As expected, $\overline {\Delta}_{1,1}$ increases and $\overline {\Delta}_{1,m} , m \neq1$ decreases as $N$ increases.
%
%
%
%
\vspace{-0mm}
\subsection{Determining the Beam Corresponding to \SURx}\label{SubSe3}
%
%
When the channel is sensed idle, \SUTx ~leaves {\it{channel sensing and monitoring phase}} and enters {\it{channel training phase}}. During this phase, \SUTx ~sends pilot symbols over all beams to enable channel training and estimation at \SURx. Using the  received training signal, \SURx ~estimates the channel gains  $\nu_m = |\chi_m|^2$  corresponding to all sectors and determines the strongest channel $\nu^*=\max \{\nu_m \}$ among all beams and the corresponding beam index $m_\text{SR}^* = {{ \arg\max}} \{ \nu_{m} \}$. For example, in Fig. \ref{i_SR}, we have $m_\text{SR}^*=2$, i.e., the second beam has the largest channel gain. \SURx ~employs an $n_b$-bit quantizer, with quantization thresholds $\{\mu_k\}_{k=0}^{N_b}$ and quantization intervals $\{\mathcal{I}_k\}_{k=0}^{N_b}$, to quantize $\nu^*$ and to find the quantization interval to which $\nu^*$ belongs to. Then, \SURx ~feeds back $ m_\text{SR}^*$  as well as the $n_b$-bit representation of the index of the quantization interval to which $\nu^*$ belongs, over the feedback link to \SUTx.
Let $\Psi_i = \Pr \{m^*_\text{SR}=i \}$ denote the probability that $m_\text{SR}^*=i$. To characterize $\Psi_i$ we need to find the CDF and pdf of $\nu^*$, denoted as $F_{\nu^*}(\cdot)$ and $f_{\nu^*}(\cdot)$, respectively.  Note that given our assumptions, $\nu_m$'s are independent across sectors, however, not necessarily identically distributed. Therefore, $F_{\nu^*}(x)$  can be written as
\vspace{-2mm}
\begin{equation}\label{F_sc}
F_{\nu^*} (x ) = \prod_{m=1}^{M}  F_{\nu_m} (x ),
\vspace{-1.5mm}
\end{equation}
%
\noindent where $F_{\nu_m} (x )=1-e^{\frac{-x}{\delta_{m}}}$. After simplification, \eqref{F_sc} can be written as
\vspace{-2mm}
\begin{equation}\label{F_sc1}
F_{\nu^*}(x) =1+ \sum_{m=1}^{M} (-1)^{m} \sum\nolimits_{m} \exp \left(-x A_{j_1:j_m} \right), 
\vspace{-2mm}
\end{equation}
%
\noindent where 
\vspace{-2.5mm}
\begin{equation*}
A_{j_1:j_m} \! = \! \sum_{i=1}^{m} \frac{1}{\delta_{j_i}}, ~~~~~~~~~\sum\nolimits_{m} \!\! = \!\! \sum_{j_1=1}^{M-m+1}~ \sum_{j_2=j_1+1}^{M-m+2} \! \cdots \!\! \sum_{j_m =j_{m-1}+1}^{M}\!\!\!\!.
\vspace{-1mm}
\end{equation*}
%
%
\noindent From the CDF in \eqref{F_sc1}, we can find the pdf
\vspace{-1.5mm}
\begin{equation}
f_{\nu^* }(x ) = \sum_{m=1}^{M} (-1)^{m+1} \sum\nolimits_{m}  A_{j_1:j_m} \exp \left(-x A_{j_1:j_m} \right). 
\vspace{-1mm}
\end{equation}
%
Similar to section \ref{PU_Beam_Selection}, we can express $\Psi_i$ as the following
\vspace{-1.5mm}
\begin{equation}\label{Prob_i}
\Psi_i = \Pr \big \{m^*_\text{SR} = i \big \} = \int_{0}^{\infty}  f_{\nu_{i}} (y)  \prod_{\underset{m \neq i} {m=1} }^{M} \! F_{\nu_{m}} (y) ~dy.
\vspace{-2mm}
\end{equation}
%
\noindent Without loss of generality, suppose $i=1$. After some mathematical simplification, $\Psi_1$ can be expressed as
\vspace{-1mm}
\begin{equation}
\Psi_1 = \Pr \big \{m_\text{SR}^*  =1 \big \}  = 1+ \sum_{ {m=1} }^{M-1} (-1)^{m} \sum\nolimits_{m}^{\prime} \frac{1}{1 + \delta_{1} B_{j_1:j_m} },
\vspace{-1.5mm}
\end{equation}
%
where
\vspace{-2.5mm}
\begin{equation*}
B_{j_1:j_m} \!\! = \! \sum_{i=1}^{m} \frac{1}{ \delta_{(1+j_i)}}, ~~~~~~\sum\nolimits_{m}^{\prime} \!\! = \!\! \sum_{j_1=1}^{M-m}~ \sum_{j_2=j_1+1}^{M-m+1} \! \cdots \!\!\! \sum_{j_m =j_{m-1}+1}^{M-1}\!\!\!\!. 
\end{equation*}
%
%
\section{Formalizing and Solving \eqref{Prob1}}\label{Sect4}
After {\it{channel training phase}}, \SUTx ~enters {\it{data transmission phase}}. Going through the previous two phases, at this point \SUTx ~knows the beam indices $m_\text{PU}^*$, $m_\text{SR}^*$ as well as the index of quantization interval to which the largest channel gain $\nu^*$ belongs to. Knowing the quantization interval index, \SUTx ~infers the quantized value of $\nu^*$ and adopts its discrete power level accordingly. For instance, if $\nu^* \in \mathcal{I}_k$ then the quantized $\nu^*$ is $\mu_k$ and the associated discrete power level is $P_k$. From a system-level design perspective, one can optimize the quantization thresholds $\mu_k$'s and the associated discrete power levels $P_k$'s, such that the constrained capacity is maximized. Furthermore, the capacity expression itself and the power of interference signal imposed on PU during this phase depend on the accuracy of the energy-based binary detector in Section \ref{GLRT_SS}, in a way that increasing the detector accuracy has a positive effect on lowering the interference power and a negative impact on enhancing the capacity itself. This implies that an optimal $T_\text{sen}$ can exist that maximizes the constrained capacity. In the following we express $C_{0,0}$ and $C_{1,0}$ in terms of the optimization variables $\{\mu_k, P_k\}_{k=1}^{N_b}$ and we find the term $\mathbb{E} \{ p( \kappa_{\text{SR}}^* \! - \! \kappa_{\text{PU}}^* )  \}$ in \eqref{Iav01} using the analysis we have conducted in sections \ref{PU_Beam_Selection} and \ref{SubSe3}. We modify the objective function and the constrains in terms of the optimization variables in Section \ref{Formal_A}. Then, we provide our solution to the problem in Section \ref{Solving}. 
%
%
\vspace{-1mm}
\subsection{Formalizing \eqref{Prob1} with Modified Objective Function and Constraints}\label{Formal_A} 
\vspace{-1mm}
Starting with the continuous valued $\nu^*$ and its corresponding continuous valued transmit power $P(\nu^*)$, we can write the expressions for the instantaneous capacity $C_{0,0}$ and $C_{1,0}$ in \eqref{C_Ergodic} as \cite{ICASSPpaper}
%
%
%
%
\vspace{-1.5mm}
\begin{equation}\label{C00C10}
C_{0,0} \! = \! \log_2 \! \left( \! 1\!+\!\frac{ \nu^* \!P(\nu^*)}{\sigma^2_\text{w}} \! \right)\!, ~~~C_{1,0} \!= \!\log_2\! \left(\!1\!+\!\frac{ \nu^* \!P(\nu^*)}{\sigma^2_\text{w}\!+\!P_\text{p}  g_\text{sp} {\color{red} } } \! \right)\!. 
\vspace{-1.5mm}
\end{equation}
%
%
\noindent Since SUs and PU cannot cooperate, \SUTx ~cannot estimate the channel gain $g_\text{sp}$ and thus $C_{1,0}$ cannot be directly maximized at \SUTx. Instead, we consider a lower bound on its average over $g_\text{sp}$, denoted as $\mathbb{E}_{g_\text{sp}}  \{C_{1,0} \}$.
Using the Jensen's inequality \cite{Cover}, the lower bound on $\mathbb{E}_{g_\text{sp}} \! \left\{  C_{1,0} \right \}$ 
becomes
\vspace{-1.5mm}
\begin{equation}
\mathbb{E}_{g_\text{sp}}  \left\{  C_{1,0} \right \} \geq  \log_2 \left(1+\frac{ \nu^* P(\nu^*)}{\sigma^2_\text{w}+ \sigma_\text{p}^2 }\right) = C_{1,0}^\text{LB}
\vspace{-1.5mm}
\end{equation}
%
where $\sigma^2_\text{p}  = P_\text{p} \mathbb{E} \{g_\text{sp}\}= P_\text{p}  \gamma_\text{sp}$. Let  $C^\text{LB}= D_t \mathbb{E}_{\nu^*} \! \big \{ \alpha_{0} C_{0,0} + \beta_0 C_{1,0}^\text{LB}  \big \}$ where $C^\text{LB}$ is the lower bound on $C$ in \eqref{C_Ergodic}. From now on, we focus on $C^\text{LB}$. Let $R_{0,0}^{(k)}$ and $R_{1,0}^{(k)}$ denote the discrete transmission rates when the quantization interval index of $\nu^*$ is $k$, i.e., $\nu^* \in \mathcal{I}_k$, quantized $\nu^*$ is $\mu_k$, and discrete power level is $P_k$. From \eqref{C00C10} we have
\vspace{-1mm}
\begin{equation}
R_{0,0}^{(k)} \! = \! \log_2 \! \left(\!1\! +\! \frac{\mu_k  P_k}{\sigma^2_\text{w}} \! \right), ~~~~~~R_{1,0}^{(k)} \! = \! \log_2 \! \left( \!1 \!+ \! \frac{\mu_k  P_k}{\sigma^2_\text{w} \! + \! \sigma^2_\text{p} } \! \right) \!. 
\vspace{-1mm}
\end{equation}
%
%
Recall that the probability of quantized $\nu^*$ being in the interval $\mathcal{I}_k$ is equal to $F_{\nu^*}(\mu_{k+1}) \! - \! F_{\nu^*}(\mu_{k})$. By averaging over all possible quantization intervals, we can rewrite $C^\text{LB}$ in terms of the discrete transmission rates as the following:
\vspace{-1.5mm}
\begin{equation}\label{Cavg}
C^\text{LB}  =  D_t \sum_{k=1}^{N_b} \! \left ( \alpha_0 R_{0,0}^{(k)} \!+\! \beta_0 R_{1,0}^{(k)} \right )  \Big [ F_{\nu^* }(\mu_{k+1}) - F_{\nu^* }(\mu_{k})  \Big ]. 
\vspace{-1mm}
\end{equation}
%
Next, we focus on the constraint in \eqref{Iav01} and find the term $\mathbb{E} \{ p( \kappa_{\text{SR}}^* \! - \! \kappa_{\text{PU}}^* )  \}$. Using the average probabilities derived  in \eqref{DeltaAverage} and \eqref{Prob_i} we have 
\vspace{-1.5mm}
\begin{equation}\label{ExpecP}
 \mathbb{E} \big \{ p ( \kappa_{ \text{SR}}^*  - \kappa_{\text{PU}}^* ) \big \}  = \sum_{j=1}^{M} \sum_{i=1}^{M} \Psi_ j ~\overline{\Delta}_{m_\text{PU}^*, i} ~ p ( \kappa_{j}  - \kappa_{i} ). 
\vspace{-1mm}
\end{equation}
%
Then, the constraint in \eqref{Iav01} can be written as 
\vspace{-1.5mm}
\begin{equation}\label{Iav}
D_t b_0 \sum_{k=1}^{N_b} P_k \Big [ F_{\nu^* }(\mu_{k+1}) \! - \! F_{\nu^* }(\mu_{k})  \Big ] \leq \Ibar, 
\vspace{-1.5mm}
\end{equation}
%
\noindent where $b_0$ is
\vspace{-2.5mm}
\begin{equation}\label{b0}
b_0 = \beta_0 \gamma_{\text{sp}} \sum_{j=1}^{M} \sum_{i=1}^{M} \Psi_j  ~\overline{\Delta}_{m_\text{PU}^*, i} ~ p ( \kappa_{j}  - \kappa_{i} ).
\vspace{-1.5mm}
\end{equation}
%
We end this section with the statement of the constrained optimization problem we solve. In Section \ref{Solving} we solve  the following constrained optimization problem
\vspace{-1mm}
\leqnomode
\begin{align}\tag{P2}\label{Prob2}
~~~~~{\underset { T_\text{sen}, \{\mu_k , P_k \} _{k=1}^{N_b}}  {\text{Maximize}} }  ~C^\text{LB} \! = \! D_t \! \sum_{k=1}^{N_b} \! & \left ( \alpha_0 R_{0,0}^{(k)} \!+\! \beta_0 R_{1,0}^{(k)} \right ) \\
& \times \Big [ F_{\nu^* }(\mu_{k+1}) - F_{\nu^* }(\mu_{k})  \Big ]  \nonumber
\end{align}
\vspace{-6mm}
\begin{align}
\begin{array}{ll}
\text{s.t.:}  ~~& 0 < T_\text{sen} < (T_\text{f}  \!-\!T_\text{train}), \nonumber \\
 &  0 < \mu_1< \ldots < \mu_{N_b} < \infty,  \nonumber \\
 &   ~P_k > 0 ~\forall k, \nonumber \\
 & \eqref{Iav}  ~\text{and} ~ \eqref{Pav01} ~\text{are satisfied.} \nonumber
 \end{array}
\end{align}
\reqnomode
%
{\color{blue} It is worth mentioning that \eqref{Prob2} includes the special case where the locations (orientations) of PU and \SURx ~are such that they belong to the same beam,  with respect to \SUTx. First, suppose $m^*_\text{PU}\!=\!m^*_\text{SR}$. In this case, the interference imposed on PU increases and \SUTx ~uses a small transmit power level $P_k$, such that the average interference constraint in \eqref{Iav} is satisfied. Next, suppose $m^*_\text{PU}\! \neq \!m^*_\text{SR}$. In this case \SUTx ~uses a larger $P_k$, compared with the case where $m^*_\text{PU}\!=\!m^*_\text{SR}$ (because \SUTx ~wrongly assumes that PU and \SURx ~lie in two different beams/sectors). Although the instantaneous interference in this case becomes larger (compared with the case where $m^*_\text{PU}\!=\!m^*_\text{SR}$), the average interference constraint in \eqref{Iav} is still satisfied.}
%
%
\vspace{-1.5mm}
\subsection{Solving \eqref{Prob2}}\label{Solving}
\vspace{-0.5mm}
\par We note that \eqref{Prob2} is a non-convex problem and can be solved using exhaustive search, which can be computationally expensive. Therefore we develop an iterative suboptimal algorithm with a much less computational complexity,  to find the local optimal solution using the Lagrangian method. The Lagrangian is 
\vspace{-2mm}
\begin{align}\label{Lagrang}
{\cal L}  = & -  D_t \sum_{k=1}^{N_b} \big ( \alpha_0 R_{0,0}^{(k)} + \beta_0 R_{1,0}^{(k)} \big ) \Big [ F_{\nu^* }(\mu_{k+1}) \! - \! F_{\nu^* }(\mu_{k})  \Big ]  \nonumber \\
 & +  \lambda \Big ( \! D_t \pioh  \sum_{k=1}^{N_b} P_k  \Big [ F_{\nu^* }(\mu_{k+1}) \! - \! F_{\nu^* }(\mu_{k})  \Big ] \! - \! \Pbar \! \Big ) \nonumber \\
 & +   \vartheta \Big ( \! D_t b_0 \sum_{k=1}^{N_b} P_k  \Big [ F_{\nu^* }(\mu_{k+1}) \! - \! F_{\nu^* }(\mu_{k})  \Big ] \! - \! \Ibar \! \Big )
\end{align}
%
%
%
\noindent where $\lambda$ and $\vartheta$ are the nonnegative Lagrange multipliers, associated with the  average transmit power and interference  constraints, 
respectively. {\color{blue}The Lagrangian multipliers can be obtained using the subgradient method}. Our iterative algorithm is based on the block coordinate descent algorithm (BCDA)  which relies on the following principle: all variables expect one are assumed to be fixed and the optimal variable that minimizes \eqref{Lagrang} is found. This process is iterated for all the variables until the final solution is reached. Convergence is achieved if there exists a single solution that minimizes \eqref{Lagrang} at each iteration \cite{Alouini}. To apply the principle of BCDA algorithm in our problem, we consider the following. Assuming fixed $\mu_k$'s and $T_\text{sen}$, the problem \eqref{Prob2} becomes convex with respect to $P_k$. Therefore, the optimal $P_k$'s that minimize \eqref{Lagrang} are the solutions to the Karush-Kuhn-Tucker (KKT) optimality necessary and sufficient conditions 
\vspace{-2mm}
\begin{align}
P_k  = & \left [ \frac{F_k + \sqrt{\Upsilon_k}}{2}  \right ] ^{+},  ~~~~~~ \text{for} ~k=1, 2, \ldots, N_b \label{Pi_All_Links} \nonumber \\
 F_k & = \frac{\pioh}{\ln(2) \left (\lambda \pioh \! + \! \vartheta b_0 \right )} -  \frac{2\sigma^2_\text{w} \! + \! \sigma^2_\text{p}}{\mu_k}, \nonumber \\
\Upsilon_k = F_k^2  & \! - \! \frac{4}{\mu_k} \! \left ( \frac{\sigma^2_\text{w} (\sigma^2_\text{w} \! + \!  \sigma^2_\text{p})}{\mu_k} \! - \! \frac{ \pioh \sigma_\text{w}^2 \! + \! \beta_0 \sigma^2_\text{p}}{ \ln(2) \left (\lambda \pioh + \vartheta b_0 \right )}    \right ), 
\end{align}
%
%
where $[x]^+ =\max(x,0)$. On the other hand, assuming fixed $P_k$'s and $T_\text{sen}$, the optimal $\mu_k$'s that minimize \eqref{Lagrang} are the solutions to $\partial {\cal L}/\partial \mu_k = 0$ for $k=1, \ldots, N_b$, which is the first derivative of ${\cal L}$ with respect to $\mu_k$. Setting $\partial {\cal L}/\partial \mu_k = 0$ we reach \eqref{Fmui}. 
Note the values of $\lambda$ and $\vartheta$ in \eqref{Pi_All_Links} and \eqref{Fmui} are obtained by applying the constraints given in \eqref{Iav} and \eqref{Pav01}. Recall that $\mu_0=0$ and $\mu_{N_b+1}=\infty$ and hence $F_{\nu^*}(\mu_0)=0$ and $F_{\nu^*}(\mu_{N_b+1})=1$. 
\par We are now ready to state our iterative algorithm to find the local optimal solution of \eqref{Prob2}. In the first step, let $T_\text{sen}$ be a value in the interval $(0, T_\text{f}-T_\text{train})$. We initiate $\mu_1>0$ and find $P_1$ using \eqref{Pi_All_Links}. Having $P_1, P_0=0$ and $\mu_1$ we obtain $\mu_2$ using \eqref{Fmui}. We repeat this and iterate between \eqref{Pi_All_Links} and \eqref{Fmui} until we find $\{P_k,\mu_k\}_{k=1}^{N_b}$. At this point, we  check whether or not $F_{\nu^*}(\mu_{N_b+1})=1$. If $F_{\nu^*}(\mu_{N_b+1})$ is less (greater) than one, we increase (decrease) the initial value of $\mu_1$ and find a new set of values for $\{P_k,\mu_k\}_{k=1}^{N_b}$ and check for the condition $F_{\nu^*}(\mu_{N_b+1})=1$. We continue changing the initial value of $\mu_1$ and finding new values for $\{P_k,\mu_k\}_{k=1}^{N_b}$ and checking for the condition $F_{\nu^*}(\mu_{N_b+1})=1$, until we find the set of values such that this condition is satisfied. In the second step, given $\{P_k,\mu_k\}_{k=1}^{N_b}$ values reached at the end of the first step, we find $T_\text{sen}$ that minimizes \eqref{Lagrang}, using search methods such as bisection method\footnote{The problem in \eqref{Prob2} can be solved offline, based on the statistical information of the channels between \SUTx-PU and \SUTx-\SURx, the number of sectors $M$, and the number of feedback bits $n_b$. In particular, given each pair $m_{ \text{PU}}^*$, $m_{ \text{SR}}^* \in \{1,...,M\}$ there is a set of optimal  solution for $T_\text{sen}$, $\{\mu_k,P_k\}_{k=1}^{N_b}$. These $M^2$ sets of solutions are available a priori at \SUTx.  Also, the $M^2$ sets of $\{\mu_k\}_{k=1}^{N_b}$ are available a priori at \SURx. During channel training phase, \SUTx ~can also send its finding $m_{ \text{PU}}^*$ to \SURx. With the knowledge of $m_{ \text{PU}}^*$ and $m_{ \text{SR}}^*$, \SURx ~would know which set of quantization thresholds to use for quantizing $\nu^*$. The idea of offline power allocation optimization with a limited feedback channel has been used before for distributed detection systems in wireless sensor networks \cite{Evans_WSN}.}. {\color{blue} A summary of our proposed iterative algorithm for solving \eqref{Prob2} is given in Algorithm \ref{Alg}.}
%
%
\begin{figure*}
\begin{equation}\label{Fmui}
F_{\nu^*} (\mu_{k+1}) = F_{\nu^*}(\mu_{k}) + \frac{f_{\nu^* }(\mu_k) \Big [ \alpha_0 \big ( R_{0,0}^{(k)} - R_{0,0}^{(k-1)} \big )+ \beta_0 \big ( R_{1,0}^{(k)} -R_{1,0}^{(k-1)} \big ) - ( \lambda \pioh +  \vartheta b_0)( P_k - P_{k-1} ) \Big ]}{ \frac{P_k}{\ln(2)} \left ( \frac{ \alpha_0}{\sigma_\text{w}^2 + \mu_k P_k } + \frac{ \beta_0}{\sigma_\text{w}^2 + \sigma_p^2 + \mu_k P_k }  \right )}
\end{equation}
\hrulefill
\vspace{-2mm}
\end{figure*}
%
%
\begin{algorithm}[!t]
\small
{\color{blue}
\caption{Our proposed iterative algorithm for solving \eqref{Prob2}}
\label{Alg}
~1: Initialize $T_\text{sen} \in \big (0, T_\text{f}-T_\text{train} \big )$, $\mu_1$, $\lambda$, $\vartheta$. \\
~2: Set $P_0 = 0$. \\
~3: \textbf{repeat} \\
~4: $\quad$ \textbf{repeat} \\
~5: $\quad~\quad$ Find $P_1$  using \eqref{Pi_All_Links}. \\
~6: $\quad~\quad$ \textbf{for} $k = 2 : N_b$ \\
~7: $\quad~\quad~\quad$ Having $P_0, \ldots, P_{k-1}$, obtain $\mu_k$ using \eqref{Fmui}.\\
~8: $\quad~\quad~\quad$ Having $\mu_k$, obtain $P_k$ using \eqref{Pi_All_Links}.\\
~9: $\quad~\quad$ \textbf{end}\\
10: $\quad~\quad$ Update $\lambda$ and $\vartheta$ using subgradient method.\\ 
11: $\quad$ \textbf{until} Constraints in \eqref{Iav} and \eqref{Pav01} are satisfied.\\
12: $\quad$ Find $F_{\nu^*}(\mu_{N_b+1})$ using \eqref{Fmui}. \\
13: $\quad$ \textbf{if} $F_{\nu^*}(\mu_{N_b+1}) < 1$\\
14: $\quad~\quad$ increase $\mu_1$. \\
15: $\quad$ \textbf{elseif} $F_{\nu^*}(\mu_{N_b+1}) > 1$\\
16: $\quad~\quad$ decrease $\mu_1$.\\
17: $\quad$ \textbf{end}  \\
18: \textbf{until} $F_{\nu^*}(\mu_{N_b+1}) = 1$\\
19:  Find $T_\text{sen}^\text{Opt}$ that maximizes $C^\text{LB}$ using bisection method. \\
}
\end{algorithm}
%
%
%
\vspace{-1.5mm}
\section{Outage and Symbol Error Probabilities }\label{Sect5}
\vspace{-0.5mm}
Two other relevant metrics to evaluate the performance of our opportunistic CR system with the ESPAR antenna at \SUTx ~are outage probability and symbol error probability (SEP), denoted as $P_\text{out}$ and $P_\text{e}$, respectively. We define $P_\text{out}$ as the probability of \SUTx ~not transmitting data due to the weak \SUTx-\SURx  ~channel. In the following, we derive closed-form expressions for $P_\text{out}$ and $P_\text{e}$, based on the solutions provided  in Section \ref{Solving}. The outage probability $P_\text{out}$ can be directly obtained using the CDF of $\nu^*$ as
\vspace{-1mm}
\begin{equation}
P_\text{out} = \Pr \big \{ P(\nu^*) \!= \! 0 \big \} = \Pr \big \{ \nu^* \!< \!\mu_1 \big \} = F_{\nu^*}(\mu_1).
\vspace{-1mm}
\end{equation}
%
For many digital modulation schemes SEP can be written as $P_\text{e} = \mathbb{E}  \left \{ Q(\sqrt{\rho ~\text{SNR}}) \right \}$ where $\rho$  is a constant parameter related to the type of modulation \cite{Evans}. Considering the noise (plus interference) imposed on \SURx ~under hypotheses $\widehat{\mathcal{H}}_0$ and $\widehat{\mathcal{H}}_1$, we can write $P_\text{e}$ as 
\vspace{-1mm}
\begin{equation}\label{Pe01}
P_\text{e}  = \alpha_0 \,\mathbb{E}  \left \{ \! Q \left (\!\sqrt{\frac{\rho \nu^* P(\nu^*)}{\sigma_\text{w}^2}}\right ) \! \right \} \!+ \! \beta_0 \,\mathbb{E}  \left \{ \! Q \left (\!\sqrt{\frac{ \rho \nu^* P(\nu^*)}{\sigma_\text{w}^2 \!+ \! \sigma_\text{p}^2}} \right ) \! \right \}.
\vspace{-1mm}
\end{equation}
%
Let focus on the expectation in the first term of \eqref{Pe01}. Since $P(\nu^*)=P_k$ when $\nu^* \in \mathcal{I}_k = [\mu_k, \mu_{k+1})$, we have
\vspace{-1mm}
\begin{align}
 \mathbb{E}   \Bigg \{ \! Q \! &\left (\! \sqrt{ \frac{\rho \nu^* P(\nu^*)}{\sigma_\text{w}^2}}\right ) \! \Bigg \}  = \int_{0}^{\infty} \!\! Q \! \left (\! \sqrt{\frac{\rho x P(x)}{\sigma_\text{w}^2}}\right ) f_{\nu^*}(x) dx \nonumber \\
 & \qquad \quad \quad=  \sum_{k=0}^{N_b} \int_{\mu_k}^{\mu_{k+1}} \!\!Q \left ( \! \sqrt{\frac{\rho x P_k}{\sigma_\text{w}^2}}\right ) f_{\nu^*}(x) dx.
\end{align}
%
Similarly, we can find the expectation in the second term of \eqref{Pe01}. Using the following equation 
\vspace{-1mm}
\begin{align}
\int_{\mu}^{\infty} \!\!\!Q(\sqrt{b x}) e^{-Ax} dx \! = \! \frac{1}{A} \!\! \left [ e^{-A\mu} Q(\sqrt{b \mu}) \!- \! \frac{ Q \big (\sqrt{\mu(2A \! + \! b)} \big ) }{\sqrt{1\!+\!\frac{2A}{b} }}  \right ],
\vspace{-2mm}
\end{align}
%
and after some manipulation, the $P_\text{e}$ in \eqref{Pe01} can be written as \eqref{Pe} 
where $V(\mu, \text{SNR}) $ is defined in \eqref{V_mu}. 
In \eqref{Pe}, $\text{SNR}_k^{(0)}$ and $\text{SNR}_k^{(1)}$ are the received SNR at \SURx ~when $\nu^* \in \mathcal{I}_k$ and the channel is sensed idle and busy, respectively, defined as
%
\begin{figure*}
\begin{equation}\label{Pe}
P_\text{e} =  \sum_{m=1}^{M} (-1)^{m+1} \sum\nolimits_{m} \sum_{k=0}^{N_b}  \bigg [ \alpha_0 \Big ( V(\mu_{k+1}, \text{SNR}_k^{(0)}) - V(\mu_{k}, \text{SNR}_k^{(0)}) \Big ) +\beta_0 \Big ( V(\mu_{k+1},  \text{SNR}_k^{(1)})  - V(\mu_k, \text{SNR}_k^{(1)}) \Big )  \bigg ]  
\vspace{-2mm}
\end{equation}
\begin{equation}\label{V_mu}
V(\mu, \text{SNR})  =  \frac{ Q \left (\sqrt{\mu (\text{SNR}+2 A_{j_1:j_m} )} \right )}{\sqrt{1+\frac{2 A_{j_1:j_m}}{\text{SNR}}}} -  e^{- \mu A_{j_1:j_m}} Q \left ( \sqrt{\mu \text{SNR} } \right )
\vspace{-1mm}
\end{equation}
\hrulefill
\vspace{-2mm}
\end{figure*}
%
%
\vspace{-1mm}
\begin{equation}
\text{SNR}_k^{(0)} =  \frac{\rho P_k }{\sigma_\text{w}^2}, ~~~~~~~~~\text{SNR}_k^{(1)} =  \frac{\rho P_k }{\sigma_\text{w}^2+\sigma_\text{p}^2}.
\vspace{-0mm}
\end{equation}
%
%

\vspace{-1mm}
\section{Simulation Results}\label{SimResults}
\begin{figure}[!t]
\label{OmniPatternAll}
\vspace{-0mm}
\centering
	\begin{subfigure}[b]{0.25\textwidth}                
		\centering	       		
		\psfrag{Dirctional}[Bl][Bl][0.27]{ESPAR}
		\psfrag{Omnidirectional}[Bl][Bl][0.27]{Omni-directional}
		\includegraphics[width=30mm]{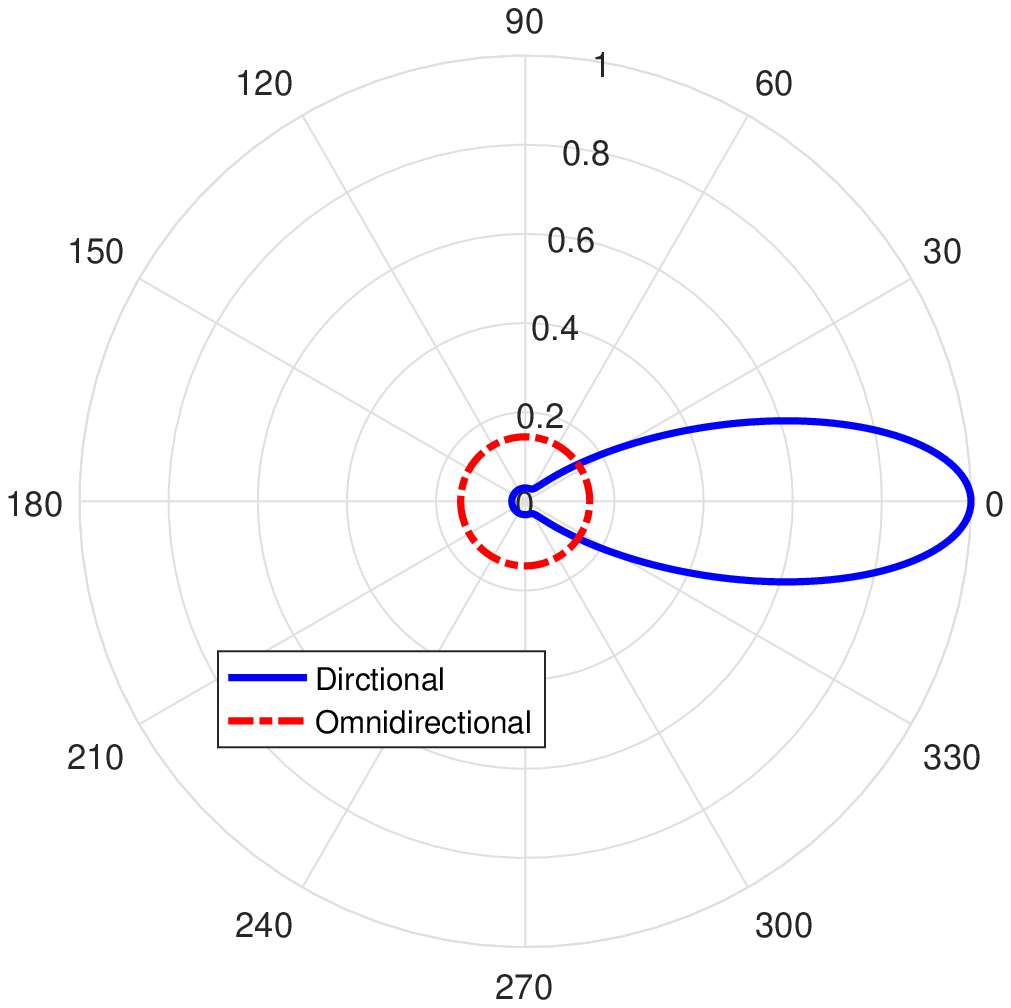}
		\vspace{-1mm}
		\caption{} 
		\label{OmniPattern}
	\end{subfigure}%
      \begin{subfigure}[b]{0.25\textwidth}
          \centering	 
     		\psfrag{Dirctional}[Bl][Bl][0.31]{ESPAR}
		\psfrag{Omnidirectional}[Bl][Bl][0.31]{Omni-directional}
		\psfrag{phi}[Bl][Bl][0.5]{$\phi$ [rad]}
		\psfrag{pphi}[Bl][Bl][0.5]{$p(\phi)$}
		\includegraphics[width=35mm]{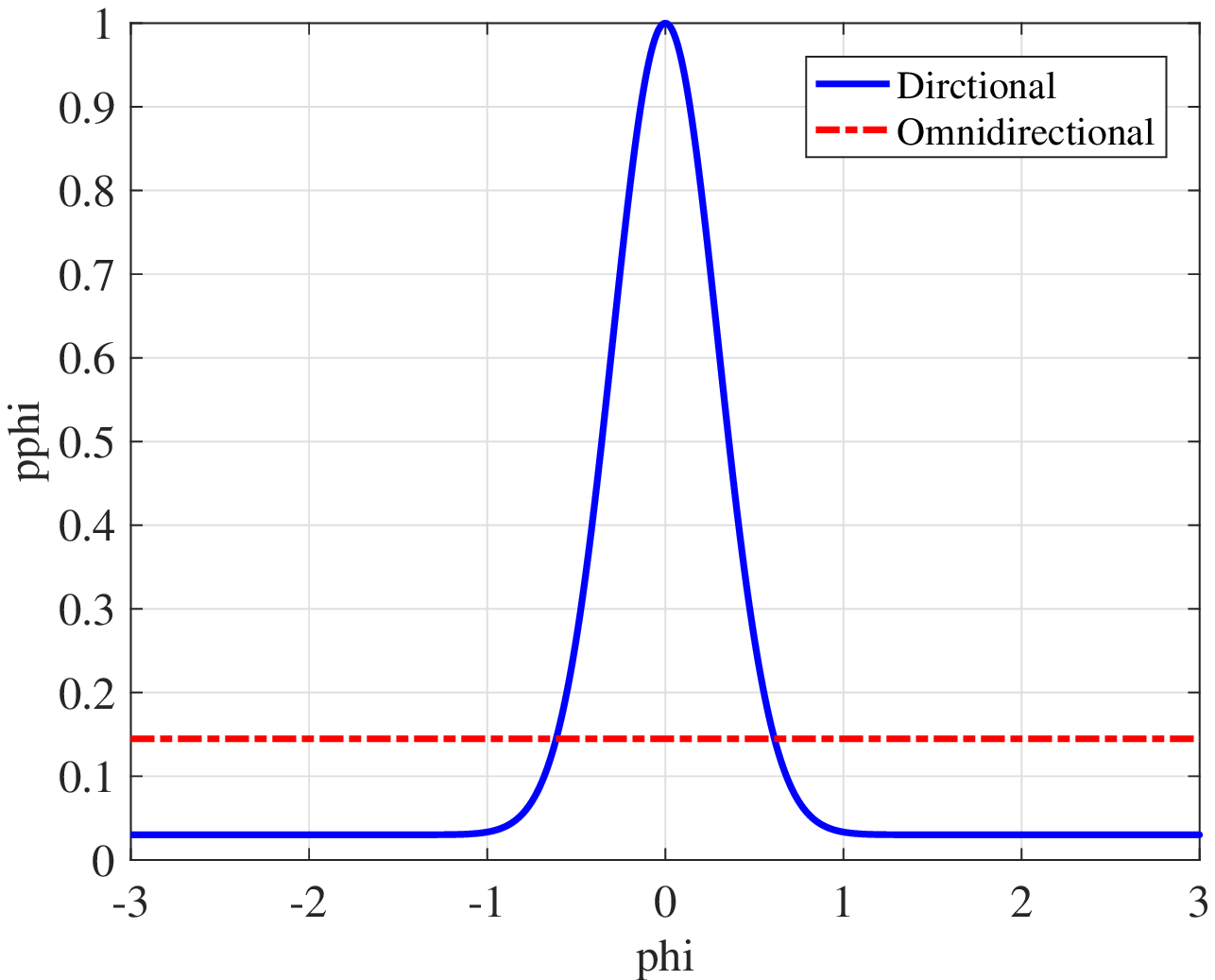}
		\vspace{-1mm}
		\caption{} 
		\label{OmniPatternCartesian}	
      \end{subfigure} \\
      \vspace{-1mm}
\caption{ {\color{blue} Parameters $A_0\!=\!0.97, A_1\!=\!0.03$, which correspond to $E_A\!=\!0.145$. For a fair comparison, we set the gain of the onmi-directional antenna $p^\text{Om}(\phi)\!=\!E_A$ for $\phi \in (-\pi,\pi)$, to ensure that the equality in \eqref{ppomni} holds true. (a) polar coordinate, (b) Cartesian coordinate.} }    
\vspace{-4mm}
\end{figure}

%
We corroborate our analysis on constrained maximization of ergodic capacity as well as outage probability and SEP derivations with Matlab simulations. To illustrate the advantage of ESPAR antennas on increasing constrained capacity, we compare the performance of our CR system with  another CR system in which \SUTx ~has an omni-directional antenna. Different from an ESPAR antenna that concentrates the electromagnetic power in specific directions (so-called sector or beam), an omni-directional antenna spreads the power equally in all angles. 
{\color{blue} To fairly compare the performance of our CR system (in which \SUTx ~has an ESPAR antenna) with the other CR system (in which \SUTx ~has an omni-directional antenna), we let $p^\text{Om}(\phi)\!=\!E_A$ for $\phi \! \in \! (-\pi,\pi)$, i.e., we set the gain of the omni-directional antenna to be $E_A$. Note that, with this setting, we have the following equality\footnote{ \color{blue} We note that comparing an ESPAR antenna with the omni-directional antenna obtained from the same ESPAR antenna is not a fair comparison  for the following reason. The omni-directional beampattern obtained from the same ESPAR antenna (when reactive loads of all parasitic elements are equal) becomes $p^\text{Om}(\phi)\!=\!A_1\!+\!A_0$ for $\phi \in (-\pi, \pi)$. Clearly, this beampattern does not satisfy the equality in \eqref{ppomni} and hence the comparison between the two CR systems is not fair. }
\vspace{-1mm}
\begin{equation}\label{ppomni}
\frac{1}{2\pi} \int_{0}^{2\pi} \!\!p(\phi)  d\phi = \frac{1}{2\pi} \int_{0}^{2\pi} \!\!p^\text{Om}(\phi)  d\phi,
\vspace{-1mm}
\end{equation}
%
Fig. \ref{OmniPattern} shows the beampatterns of omni-directional and ESPAR antennas in polar coordinate, where $A_0\!=\!0.97$, $A_1\!=\!0.03$ (corresponding to $E_A\!=\!0.145$). Note that the radius of the red beampattern is $0.145$ and the blue beampattern has the maximum value of $p(0)\!=\!A_1\!+\!A_0\! = \!1$ at angle $\phi\!=\!0$ radians. The area covered by the solid blue beampattern is equal to the area covered by the dashed red beampattern, in the sense that the equality in \eqref{ppomni} holds true. Fig. \ref{OmniPatternCartesian} plots the same beampatterns in Cartesian coordinate.} For the CR system with the omni-directional antenna at \SUTx, we consider a {\color{blue}modified} procedure for  {\it channel sensing and monitoring}, {\it channel training} and {\it data transmission} phases\footnote{\color{blue} Since the omni-directional antenna has only one beampattern, there is no beam selection corresponding to the orientations of PU and \SURx. Thus,  step 1.3 of Table  \ref{tab:Summery1} will be removed. The following steps in Table \ref{tab:Summery1} are modified: in step 2.2, \SURx ~estimates only one channel gain $\nu$, in step 2.4, \SURx ~feeds back only the $n_b$-bit representation of the index of the quantization interval to which $\nu$ belongs to \SURx, in step 3.1, \SUTx ~adapts its discrete power level $P_k$, using the information received from \SURx.}{ \color{blue}  (with respect to the description in Section \ref{ProblemStatement}) and denote the constrained capacity in \eqref{Prob2} evaluated at the optimized variables $T_\text{sen}$, $\mu_k$'s, $P_k$'s,  by $C^\text{LB,Om}_\text{Opt}$}. {\color{blue} For our CR system let} $C^\text{LB}_\text{Opt}$ denote the constrained capacity in \eqref{Prob2}, that is evaluated at the optimized variables $T_\text{sen}, \mu_k$'s, $P_k$'s. {\color{blue} Obviously, the optimized variables obtained from solving \eqref{Prob2} for omni-directional and ESPAR antennas can be different. }
%
%
\begin{table}[t!]
  \begin{center}
  {\color{blue}
  \vspace{-2mm}
    \caption{Simulation Parameters}
    \label{table2}
    \begin{footnotesize}
    \begin{tabular}{|l|l||l|l||l|l|} 
     \hline
      \textbf{Parameter} & \textbf{Value} & \textbf{Parameter} & \textbf{Value} & \textbf{Parameter} & \textbf{Value}\\
      \hline \hline
      $A_0$ & $1, 2$ & $\gamma_\text{ss}$ & $3$ & $\sigma^2_\text{w}$ & $1$ \\
      $A_1$ & $0.01$ & $\gamma, \gamma_\text{sp}$ & $1$ & $P_\text{p}$ & $1$ watts \\
      $\phi_{3\text{dB}}$ & $20 \degree$ & $\pi_1$ & $0.3$ & $T_\text{f}$ & $20$ ms\\
      $\rho$ & $4$ & ${\overline{P}}_\text{d}$ & $0.9$ & $\Ibar$ & $-6$\,dB\\
      \hline
    \end{tabular}
    \end{footnotesize}   
    }
   \end{center}    
   \vspace{-3mm}
\end{table}
%
%

%
%
\par
\begin{figure}[!t]
\vspace{-0mm}
\centering
	\begin{subfigure}[b]{0.25\textwidth}                
		\centering	
		\psfrag{ESPAR, n  =  2}[Bl][Bl][0.38]{ESPAR, $n_b=2$}
		\psfrag{ESPAR, n  =  3}[Bl][Bl][0.38]{ESPAR, $n_b=3$}
		\psfrag{ESPAR, n  =  4}[Bl][Bl][0.38]{ESPAR, $n_b=4$}
		\psfrag{ESPAR, n  =    in}[Bl][Bl][0.38]{ESPAR, $n_b=\infty$}
		\psfrag{Omni, n  =    in}[Bl][Bl][0.38]{Omni, $n_b=\infty$}		
		\psfrag{Omni, n  =  2}[Bl][Bl][0.38]{Omni, $n_b=2$}
		\psfrag{Omni, n  =  3}[Bl][Bl][0.38]{Omni, $n_b=3$}
		\psfrag{Omni, n  =  4}[Bl][Bl][0.38]{Omni, $n_b=4$}				
		\psfrag{Omni}[Bl][Bl][0.35]{Omni}	
		\psfrag{ESPAR}[Bl][Bl][0.35]{ESPAR}		
		\psfrag{Pav}[Bl][Bl][0.50]{$\Pbar$ [\text{dB}]}									
		\psfrag{Maximized Capacity}[Bl][Bl][0.50]{Maximized Capacity}
		\includegraphics[width=42mm]{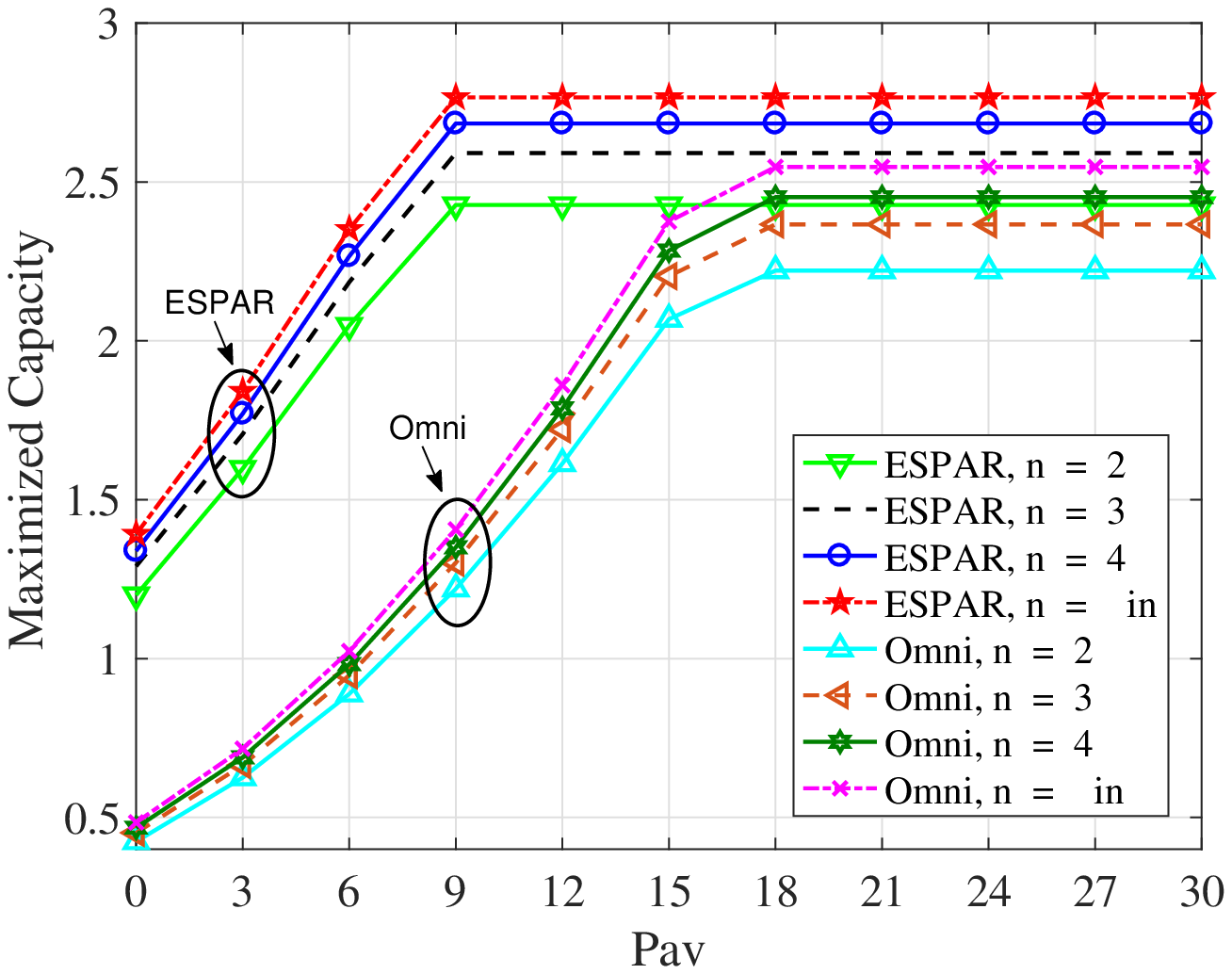}
		\vspace{-1mm}
		\caption{}
		\label{C_I6_m1}
	\end{subfigure}%
    \begin{subfigure}[b]{0.25\textwidth}
		\centering	              
		\psfrag{ESPAR, n  =  2}[Bl][Bl][0.38]{ESPAR, $n_b=2$}
		\psfrag{ESPAR, n  =  3}[Bl][Bl][0.38]{ESPAR, $n_b=3$}
		\psfrag{ESPAR, n  =  4}[Bl][Bl][0.38]{ESPAR, $n_b=4$}
		\psfrag{ESPAR, n  =    in}[Bl][Bl][0.38]{ESPAR, $n_b=\infty$}
		\psfrag{Omni, n  =    in}[Bl][Bl][0.38]{Omni, $n_b=\infty$}		
		\psfrag{Omni, n  =  2}[Bl][Bl][0.38]{Omni, $n_b=2$}
		\psfrag{Omni, n  =  3}[Bl][Bl][0.38]{Omni, $n_b=3$}
		\psfrag{Omni, n  =  4}[Bl][Bl][0.38]{Omni, $n_b=4$}						
		\psfrag{Omni}[Bl][Bl][0.35]{Omni}	
		\psfrag{ESPAR}[Bl][Bl][0.35]{ESPAR}				
		\psfrag{Pav}[Bl][Bl][0.50]{$\Pbar$ [\text{dB}]}
		\psfrag{Maximized Capacity}[Bl][Bl][0.50]{Maximized Capacity}
		\includegraphics[width=42mm]{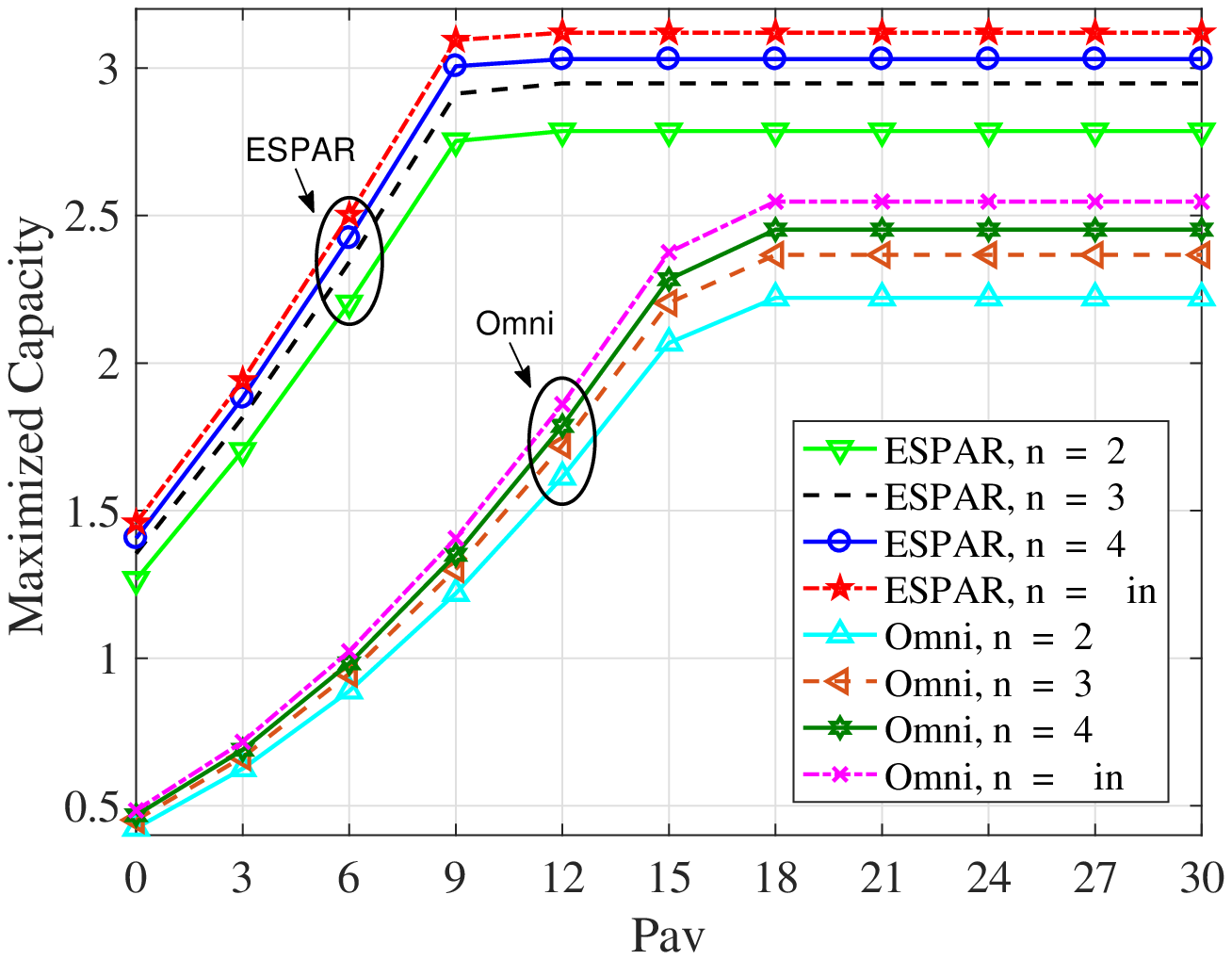}
		\vspace{-1mm}
		\caption{}
		\label{C_M12_I6_m1}
      \end{subfigure} \\
      \vspace{-1mm}
\caption{$C^\text{LB}_\text{Opt}$ {\color{blue} and  $C^\text{LB,Om}_\text{Opt}$} versus $\Pbar$ for $ m_\text{SR}^*\!=\!m_\text{PU}^*\!=\!1$ and (a) $M\!=\!8$, (b)  $M\!=\!12$. }
\vspace{-3mm}
\end{figure}

 {\color{blue} Our simulation parameters are  given in Table  \ref{table2}. First, we explore the effect of increasing the number of quantization bits $n_b$.} Fig. \ref{C_I6_m1} shows $C^\text{LB}_\text{Opt}$ {\color{blue} and $C^\text{LB,Om}_\text{Opt}$ versus $\Pbar$} for different $n_b$, when $M\!=\!8$, $m^*_\text{PU}\!=\!1$ ($\phi_\text{PU}\!=\! 12^\degree$) , $m^*_\text{SR}\!=\!1$ ($\phi_\text{SR}\!=\! 0^\degree$) and {\color{blue}$A_0\!=\!1 ,A_1 \!=\!0.01$ (corresponding to $E_A\!=\!0.127$)}. As a baseline we also plot the capacity when perfect CSI (for \SUTx-\SURx ~link) is available for both CR systems {\color{blue} (labeled as $n_b\! =\! \infty$ in the figures)}. Clearly, our CR system with the ESPAR antenna at \SUTx ~yields a higher capacity than the CR system with the omni-directional antenna at \SUTx. This figure also shows that as $n_b$ increases, $C^\text{LB}_\text{Opt}$ increases and for $n_b=4$ bits $C^\text{LB}_\text{Opt}$ is very close to the baseline capacity. To observe the impact of increasing the number of beams (the number of parasitic elements of the ESPAR antenna), Fig. \ref{C_M12_I6_m1} plots $C^\text{LB}_\text{Opt}$ {\color{blue} and $C^\text{LB,Om}_\text{Opt}$ versus $\Pbar$} {\color{blue} for different $n_b$, when} $M=12$. Comparing Figs. \ref{C_I6_m1} and \ref{C_M12_I6_m1} we observe that as $M$ increases a higher capacity can be achieved. 
\par
\begin{figure}[!t]
\vspace{-0mm}
\centering
	\begin{subfigure}[b]{0.25\textwidth}                
 		\centering     		
		\psfrag{ESPAR, n  =  2}[Bl][Bl][0.33]{ESPAR, $n_b=2$}
		\psfrag{ESPAR, n  =  3}[Bl][Bl][0.33]{ESPAR, $n_b=3$}
		\psfrag{ESPAR, n  =  4}[Bl][Bl][0.33]{ESPAR, $n_b=4$}
		\psfrag{ESPAR, n  =    in}[Bl][Bl][0.33]{ESPAR, $n_b=\infty$}
		\psfrag{Omni, n  =    in}[Bl][Bl][0.33]{Omni, $n_b=\infty$}		
		\psfrag{Omni, n  =  2}[Bl][Bl][0.33]{Omni, $n_b=2$}
		\psfrag{Omni, n  =  3}[Bl][Bl][0.33]{Omni, $n_b=3$}
		\psfrag{Omni, n  =  4}[Bl][Bl][0.33]{Omni, $n_b=4$}				
		\psfrag{Omni}[Bl][Bl][0.35]{Omni}	
		\psfrag{ESPAR}[Bl][Bl][0.35]{ESPAR}		
		\psfrag{Pav}[Bl][Bl][0.50]{$\Pbar$ [\text{dB}]}						
		\psfrag{Maximized Capacity}[Bl][Bl][0.50]{Maximized Capacity}
		\includegraphics[width=42mm]{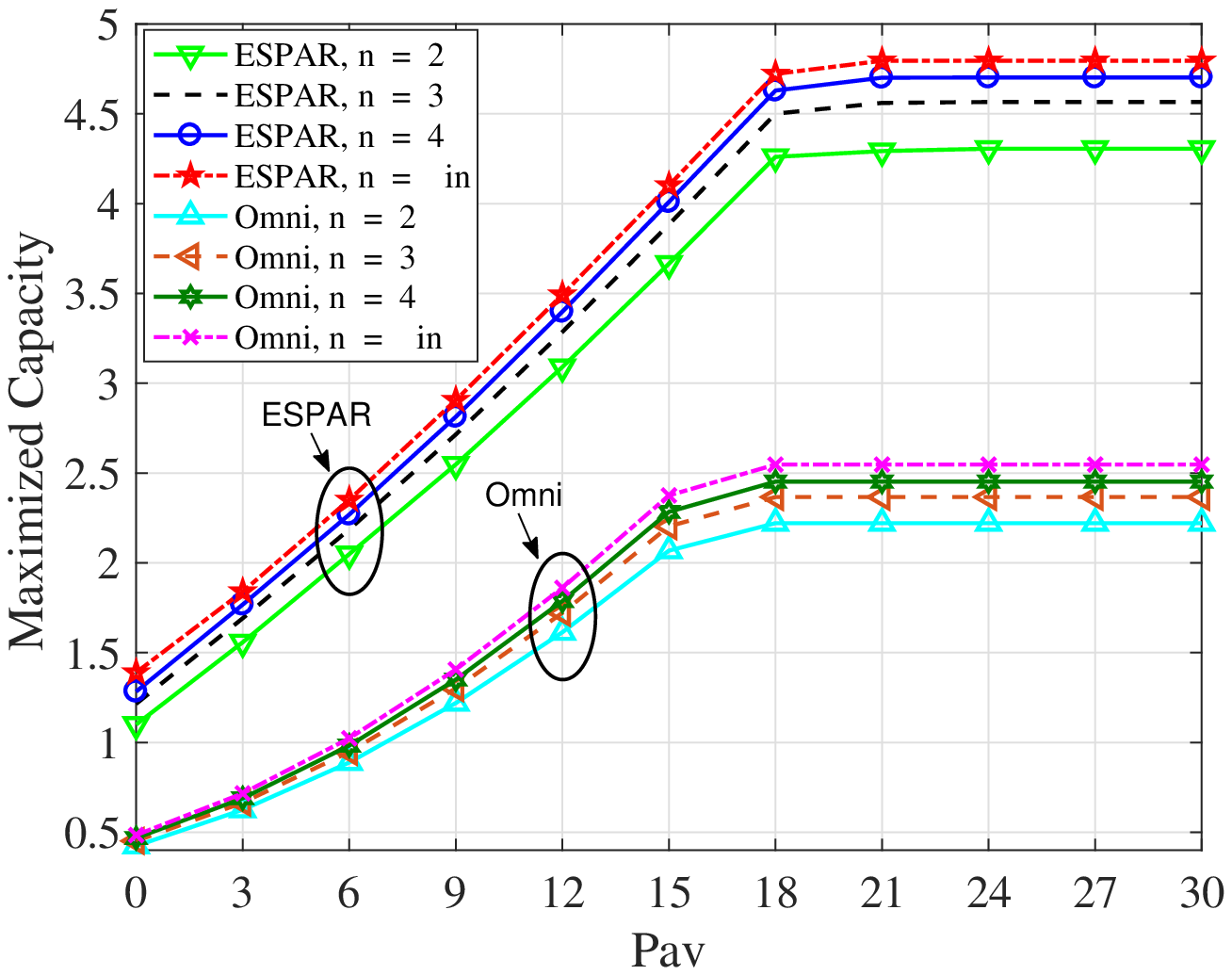}
		\vspace{-1mm}
		\caption{}
		\label{C_I6_m2}
	\end{subfigure}%
      \begin{subfigure}[b]{0.25\textwidth}
		\centering	              		
		\psfrag{ESPAR, n  =  2}[Bl][Bl][0.38]{ESPAR, $n_b=2$}
		\psfrag{ESPAR, n  =  3}[Bl][Bl][0.38]{ESPAR, $n_b=3$}
		\psfrag{ESPAR, n  =  4}[Bl][Bl][0.38]{ESPAR, $n_b=4$}
		\psfrag{ESPAR, n  =    in}[Bl][Bl][0.38]{ESPAR, $n_b=\infty$}
		\psfrag{Omni, n  =    in}[Bl][Bl][0.38]{Omni, $n_b=\infty$}		
		\psfrag{Omni, n  =  2}[Bl][Bl][0.38]{Omni, $n_b=2$}
		\psfrag{Omni, n  =  3}[Bl][Bl][0.38]{Omni, $n_b=3$}
		\psfrag{Omni, n  =  4}[Bl][Bl][0.38]{Omni, $n_b=4$}				
		\psfrag{Omni}[Bl][Bl][0.35]{Omni}	
		\psfrag{ESPAR}[Bl][Bl][0.35]{ESPAR}		
		\psfrag{Pav}[Bl][Bl][0.50]{$\Pbar$ [\text{dB}]}					
		\psfrag{Maximized Capacity}[Bl][Bl][0.50]{Maximized Capacity}
		\includegraphics[width=42mm]{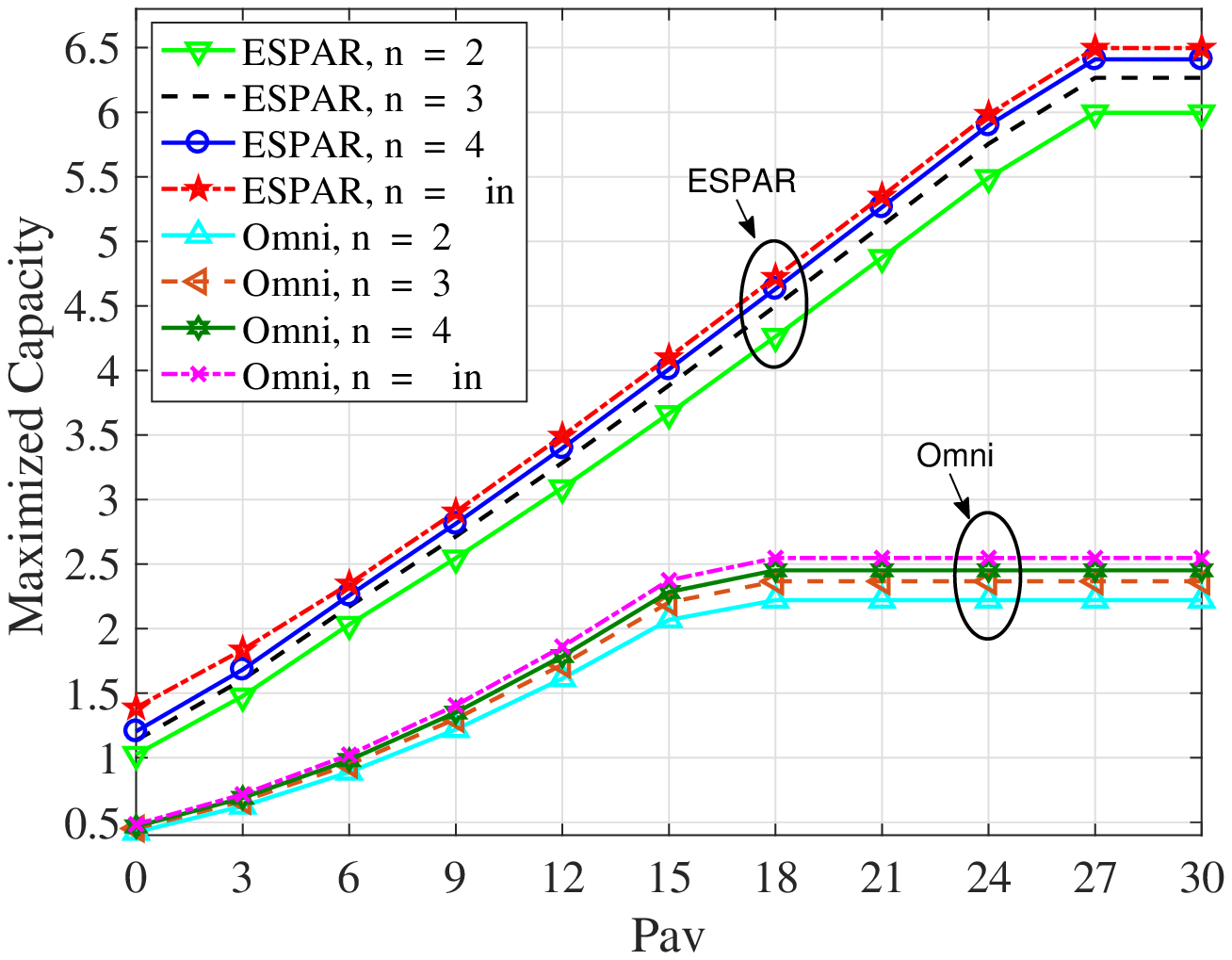}
		\vspace{-1mm}
		\caption{}
		\label{C_I6_m3}
      \end{subfigure} \\
      \vspace{-1mm}
\caption{$C^\text{LB}_\text{Opt}$ {\color{blue} and  $C^\text{LB,Om}_\text{Opt}$} versus $\Pbar$ for $M\!=\!8$,  $m_\text{SR}^*\!=\!1$ and (a) $m_\text{PU}^*\!=\!2$, (b) $m_\text{PU}^*\!=\!3$.}
\vspace{-4mm}
\end{figure}

\begin{figure}[!t]
\vspace{-0mm}
\centering
	\begin{subfigure}[b]{0.25\textwidth}                
		\centering									
		\psfrag{ESPAR, n  =  2}[Bl][Bl][0.38]{ESPAR, $n_b=2$}
		\psfrag{ESPAR, n  =  3}[Bl][Bl][0.38]{ESPAR, $n_b=3$}
		\psfrag{ESPAR, n  =  4}[Bl][Bl][0.38]{ESPAR, $n_b=4$}
		\psfrag{ESPAR, n  =    in}[Bl][Bl][0.38]{ESPAR, $n_b=\infty$}
		\psfrag{Omni, n  =    in}[Bl][Bl][0.38]{Omni, $n_b=\infty$}		
		\psfrag{Omni, n  =  2}[Bl][Bl][0.38]{Omni, $n_b=2$}
		\psfrag{Omni, n  =  3}[Bl][Bl][0.38]{Omni, $n_b=3$}
		\psfrag{Omni, n  =  4}[Bl][Bl][0.38]{Omni, $n_b=4$}				
		\psfrag{Omni}[Bl][Bl][0.35]{Omni}	
		\psfrag{ESPAR}[Bl][Bl][0.35]{ESPAR}		
		\psfrag{Pav}[Bl][Bl][0.50]{$\Pbar$ [\text{dB}]}													
		\psfrag{Average Maximized Capacity}[Bl][Bl][0.50]{Average Maximized Capacity}
		\includegraphics[width=42mm]{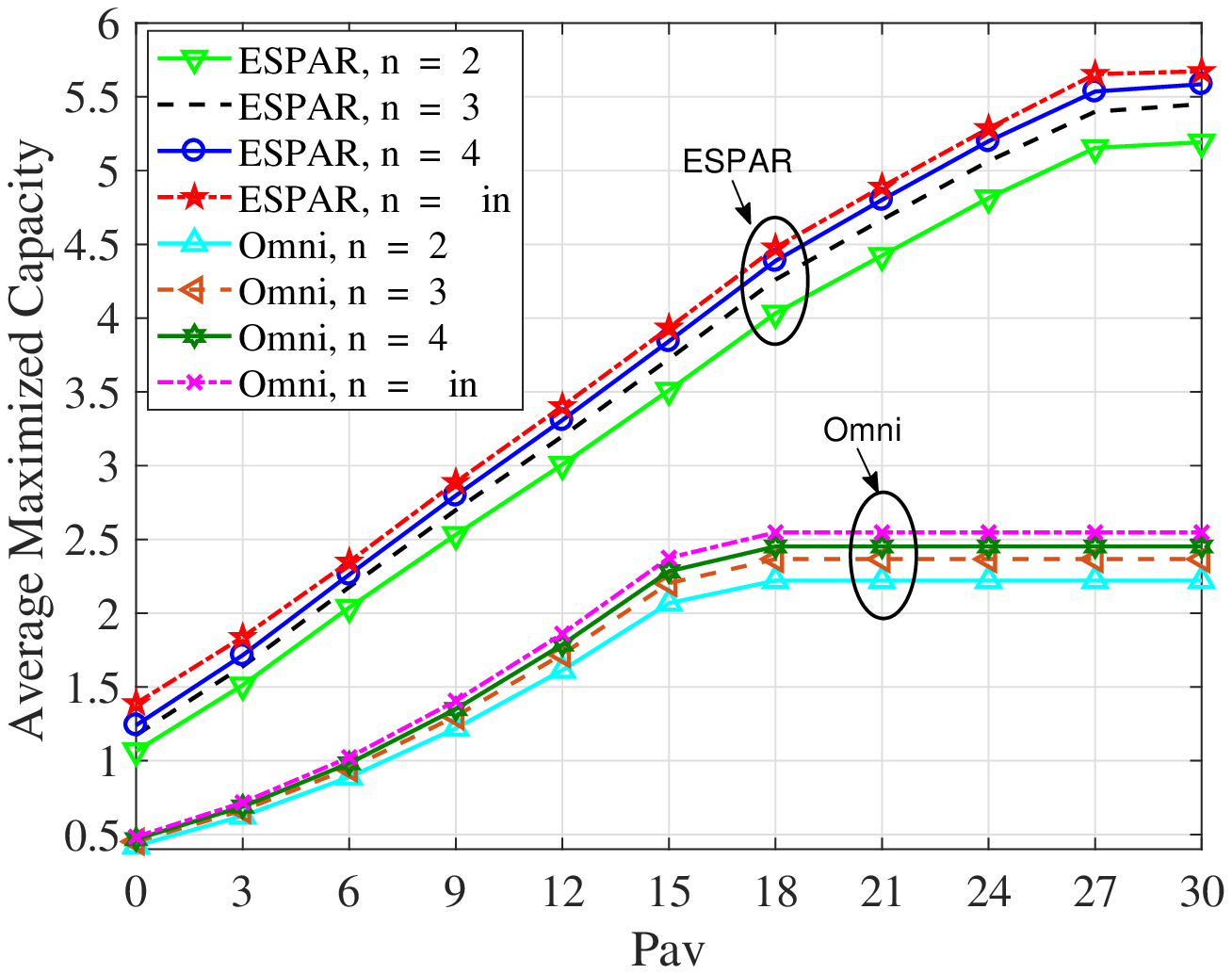}
		\vspace{-1mm}
		\caption{}
		\label{C_I6_bar}
	\end{subfigure}%
      \begin{subfigure}[b]{0.25\textwidth}
 		\centering     
		\psfrag{Iav  =  -6 dB}[Bl][Bl][0.4]{$\Ibar\!=\!-6$\,dB}
		\psfrag{Iav  =  -2 dB}[Bl][Bl][0.4]{$\Ibar\!=\!-2$\,dB}
		\psfrag{Iav  =  2 dB}[Bl][Bl][0.4]{$\Ibar\!=\!+2$\,dB}
		\psfrag{Pav}[Bl][Bl][0.60]{$\Pbar$ [\text{dB}]}
		\psfrag{Ratio}[Bl][Bl][0.6]{$\Lambda$}	  		 
		\includegraphics[width=42mm]{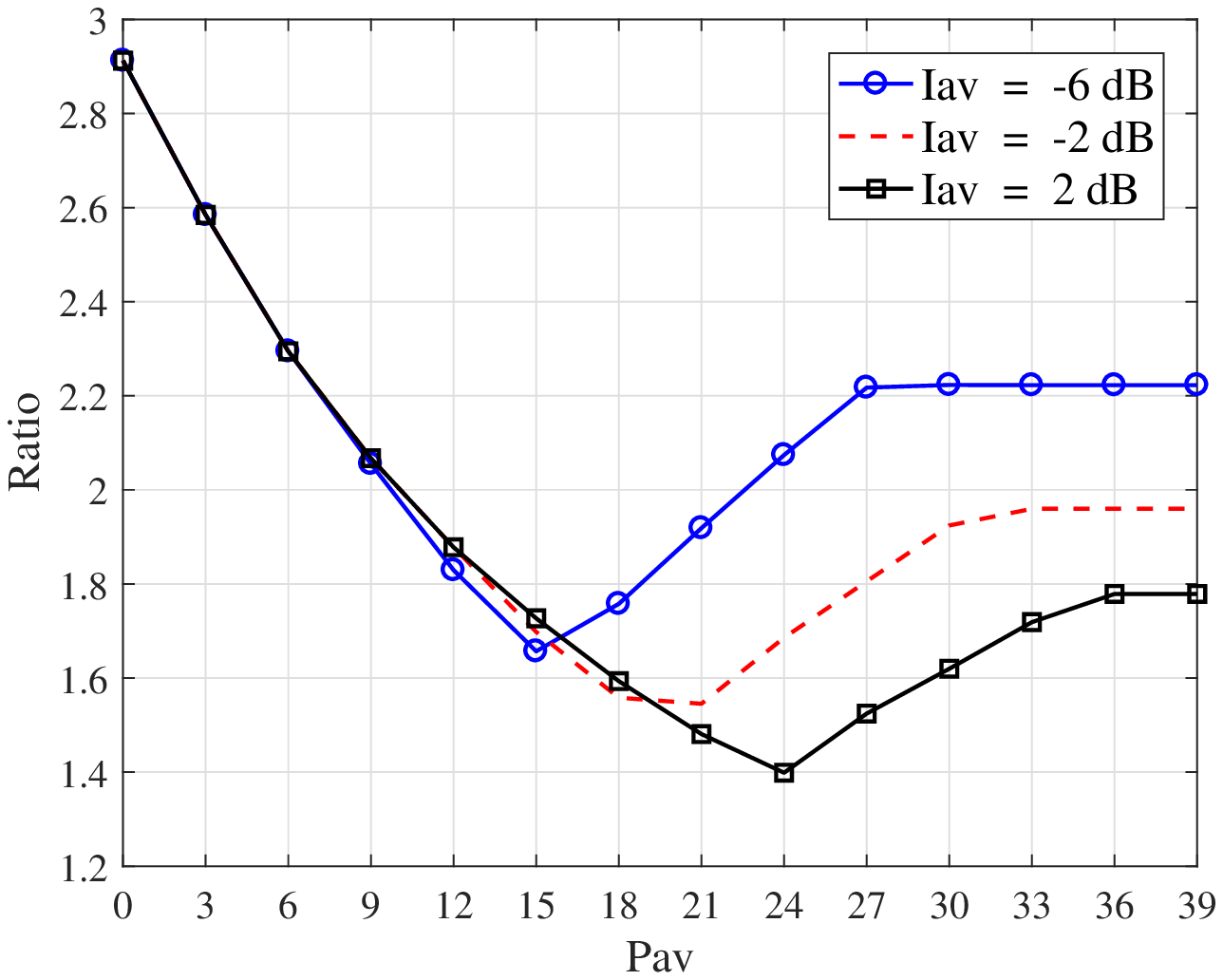}	
		\vspace{-1mm}
		\caption{}
		\label{Ratio1}      
      \end{subfigure} \\
      \vspace{-1mm}
\caption{ (a) $\overline{C^\text{LB}_\text{Opt}}$ {\color{blue}and  $C^\text{LB,Om}_\text{Opt}$} versus $\Pbar$, {\color{blue}(b) $\Lambda$ versus $\Pbar$. }}
\vspace{-6mm}
\end{figure}

\begin{figure}[!t]
\centering		
\psfrag{ESPAR,   A0=1 EA = 0.127}[Bl][Bl][0.4]{ESPAR, $A_0\!=\!1 ~(E_A\! =\! 0.127)$}
\psfrag{ESPAR,   A0=2 EA = 0.245}[Bl][Bl][0.4]{ESPAR, $A_0\!=\!2 ~(E_A\! =\! 0.245)$}
\psfrag{Omni,   A0=1 EA = 0.127}[Bl][Bl][0.4]{Omni, $A_0\!=\!1 ~(E_A\! =\! 0.127)$}		
\psfrag{Omni,   A0=2 EA = 0.245}[Bl][Bl][0.4]{Omni,  $A_0\!=\!2 ~(E_A\! =\! 0.245)$}		
\psfrag{Pav}[Bl][Bl][0.5]{$\Pbar$ [\text{dB}]}													
\psfrag{Average Maximized Capacity}[Bl][Bl][0.5]{~~Average Maximized Capacity}
\psfrag{Omni}[Bl][Bl][0.4]{Omni}	
\psfrag{ESPAR}[Bl][Bl][0.4]{ESPAR}									
\includegraphics[width=46mm]{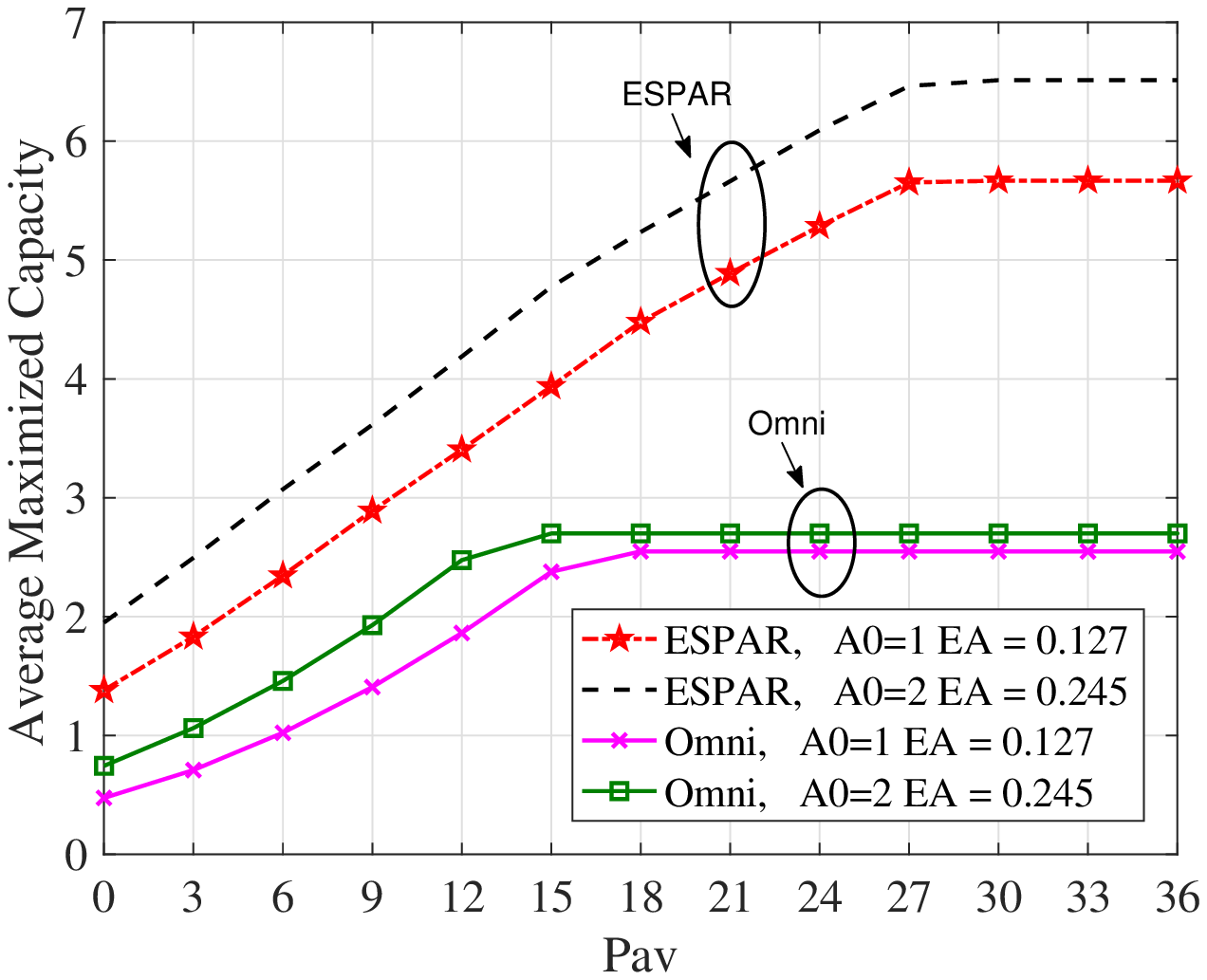}
\vspace{-1mm}
\caption{{\color{blue}$\overline{C^\text{LB}_\text{Opt}}$ and $C^\text{LB,Om}_\text{Opt}$ versus $\Pbar$.}}
\label{CPavEA}
\vspace{-2mm}
\end{figure}

 To explore the effect of changes in PU orientation, Figs. \ref{C_I6_m2} and \ref{C_I6_m3} illustrate $C^\text{LB}_\text{Opt}$ {\color{blue} and $C^\text{LB,Om}_\text{Opt}$ versus $\Pbar$} for $M\!=\!8$ when $m^*_\text{PU}\!=\!2$ and  $m^*_\text{PU}\!=\!3$, respectively (with fixed $m^*_\text{SR}\!=\!1$). Comparing Figs. \ref{C_I6_m1}, \ref{C_I6_m2}, \ref{C_I6_m3} we observe that as $m^*_\text{PU}$ becomes further away from $m^*_\text{SR}$, the imposed interference on PU from \SUTx ~decreases and \SUTx ~can transmit at a higher transmit power level, leading to an increase in $C^\text{LB}_\text{Opt}$. Note that 
{\color{blue} $C^\text{LB,Om}_\text{Opt}$} in Figs. \ref{C_I6_m1}, \ref{C_M12_I6_m1}, \ref{C_I6_m2}, \ref{C_I6_m3} are the same. 
Let $\overline{C^\text{LB}_\text{Opt}}$ denote $C^\text{LB}_\text{Opt}$ that is averaged over all possible $\phi_\text{SR}^*$ and $\phi_\text{PU}^*$.  Fig. \ref{C_I6_bar} {\color{blue} plots $\overline{C^\text{LB}_\text{Opt}}$ and $C^\text{LB,Om}_\text{Opt}$ versus $\Pbar$ for $n_b=2,3,4, \infty$. Clearly, } our CR system with the ESPAR antenna at \SUTx ~yields  a higher capacity on average, compared to the CR system with the omni-directional antenna at \SUTx.
{\color{blue} \par To quantify the capacity improvement provided with the ESPAR antenna, we define the ratio  $\Lambda = { \overline{C^\text{LB}_\text{Opt}} } / { C^\text{LB,Om}_\text{Opt} } $. Fig. \ref{Ratio1} shows $\Lambda$ versus $\Pbar$ for $\Ibar \!=\! -6, -2, 2$\,dB and $n_b\!=\! \infty$.  First, we consider how $\Lambda$ behaves as $\Pbar$ increases,  for a given $\Ibar$ value. Fig. \ref{Ratio1} shows that, as $\Pbar$ increases from zero to a certain value, $\Lambda$ decreases. As  $\Pbar$ increases beyond that certain value, $\Lambda$ increases, however, it becomes constant after $\Pbar$ reaches a certain point. For instance, given $\Ibar\!=\!-6$\,dB, $\Lambda$ decreases from $2.9$  to $1.65$, as $\Pbar$ increases from zero to $15$\,dB, it increases from $1.65$ to $2.22$, as  $\Pbar$ increases from $15$\,dB to $27$\,dB, and it becomes constant afterward. 
The reason for this behavior is that, when  $\Pbar\! \leq \!15$\,dB, the average transmit power constraint in \eqref{Pav01} is dominant for both ESPAR and omni-directional antennas. For $15\,$dB $\! \leq \! \Pbar \! \leq \! 27$\,dB, the average transmit power constraint is dominant for the ESPAR antenna and the average interference constraint in \eqref{Iav} is dominant for the omni-directional antenna. For  $\Pbar \! \geq \! 27$\,dB, the average interference constraint is dominant for both ESPAR and omni-directional antennas. 
Next, we examine how $\Lambda$ behaves as  $\Ibar$ decreases, for a given  $\Pbar$ value. Fig. \ref{Ratio1} shows that, for  $\Pbar \! \leq \! 15$\,dB $\Lambda$ does not vary much as  $\Ibar$ decreases, since the average transmit power constraint is dominant. However, this behavior changes as  $\Pbar$ increases beyond $15$\,dB, where we note $\Lambda$ increases as  $\Ibar$ decreases. Overall, we observe that the ESPAR antenna can provide a high capacity improvement ($\Lambda$ varies between $1.4$ and $2.9$ in Fig. \ref{Ratio1}),  compared with the omni-directional antenna, and the capacity improvement changes as $\Pbar$ and  $\Ibar$ vary. }
%
%
%
%
%
%
%
%
%
%
%
{\color{blue}
\par Next, we explore the influence of parameter $E_A$ defined in \eqref{EmmAb}. Fig. \ref{CPavEA} plots $\overline{C^\text{LB}_\text{Opt}}$ and $C^\text{LB,Om}_\text{Opt}$ versus $\Pbar$ for $A_1\!=\!0.01, n_b\!=\!\infty$ and two choices of $A_0$:  $A_0\!=\!1$ (corresponding to $E_A\!=\!0.127$) and $A_0\!=\!2$ (corresponding to  $E_A\!=\!0.245$). We observe that, for a given $\Pbar$ value, when we increase $A_0\!=\!1$  to $A_0\!=\!2$, the capacity enhancement  for the ESPAR antenna is higher than that of the omni-directional antenna.
To explain this observation, let  $L=A_0/A_1$ denote the ESPAR beampattern attenuation in side-lobe with respect  to its maximum value (main-lobe).
Increasing $L$ positively affects $\overline{C^\text{LB}_\text{Opt}}$ in two ways. First, the ESPAR antenna can reduce the imposed interference on PU more effectively, and hence \SUTx ~can transmit at higher power levels, without violating the average interference constraint. Second, \SUTx-\SURx ~link becomes a stronger link for data communication. 
Increasing $L$, however, affects $C^\text{LB,Om}_\text{Opt}$ differently. We note that, although increasing $L$ renders \SUTx-\SURx ~link a stronger link for data communication (positive impact), it increases the imposed interference on PU (negative impact), and hence \SUTx ~is enforced to transmit at lower power levels to satisfy the average interference constraint.}
\par
\begin{figure}[!t]
\vspace{-0mm}
\centering
	\begin{subfigure}[b]{0.25\textwidth}                
		\centering	              		
		\psfrag{ESPAR, n  =  2}[Bl][Bl][0.34]{ESPAR, $n_b=2$}
		\psfrag{ESPAR, n  =  3}[Bl][Bl][0.34]{ESPAR, $n_b=3$}
		\psfrag{ESPAR, n  =  4}[Bl][Bl][0.34]{ESPAR, $n_b=4$}
		\psfrag{ESPAR, n  =    in}[Bl][Bl][0.34]{ESPAR, $n_b=\infty$}
		\psfrag{Omni, n  =    in}[Bl][Bl][0.34]{Omni, $n_b=\infty$}		
		\psfrag{Omni, n  =  2}[Bl][Bl][0.34]{Omni, $n_b=2$}
		\psfrag{Omni, n  =  3}[Bl][Bl][0.34]{Omni, $n_b=3$}
		\psfrag{Omni, n  =  4}[Bl][Bl][0.34]{Omni, $n_b=4$}		
		\psfrag{Omni}[Bl][Bl][0.35]{Omni}	
		\psfrag{ESPAR}[Bl][Bl][0.35]{ESPAR}
		\psfrag{Pav}[Bl][Bl][0.50]{$\Pbar$ [\text{dB}]}
		\psfrag{Average Outage Probability}[Bl][Bl][0.50]{Average Outage Probability}
		\includegraphics[width=42mm]{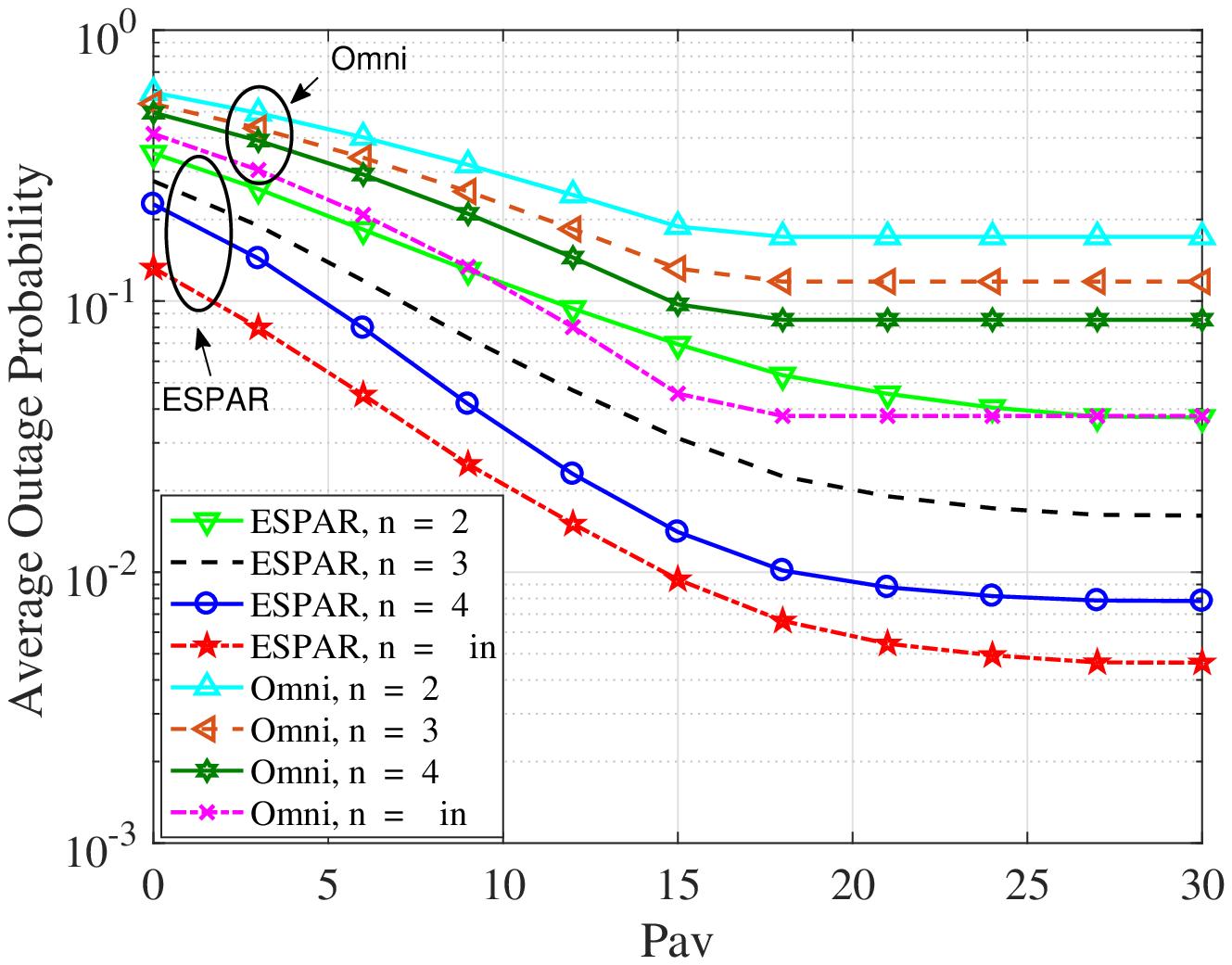}
		\vspace{-1mm}
		\caption{}
		\label{Pout_I6}			
	\end{subfigure}%
      \begin{subfigure}[b]{0.25\textwidth}
 		\centering      
		\psfrag{ESPAR, n  =  2}[Bl][Bl][0.31]{ESPAR, $n_b=2$}
		\psfrag{ESPAR, n  =  3}[Bl][Bl][0.31]{ESPAR, $n_b=3$}
		\psfrag{ESPAR, n  =  4}[Bl][Bl][0.31]{ESPAR, $n_b=4$}
		\psfrag{ESPAR, n  =    in}[Bl][Bl][0.31]{ESPAR, $n_b=\infty$}
		\psfrag{Omni, n  =    in}[Bl][Bl][0.31]{Omni, $n_b=\infty$}		
		\psfrag{Omni, n  =  2}[Bl][Bl][0.31]{Omni, $n_b=2$}
		\psfrag{Omni, n  =  3}[Bl][Bl][0.31]{Omni, $n_b=3$}
		\psfrag{Omni, n  =  4}[Bl][Bl][0.31]{Omni, $n_b=4$}				
		\psfrag{Omni}[Bl][Bl][0.35]{Omni}	
		\psfrag{ESPAR}[Bl][Bl][0.35]{ESPAR}		
		\psfrag{Pav}[Bl][Bl][0.50]{$\Pbar$ [\text{dB}]}		
		\psfrag{Average SER}[Bl][Bl][0.50]{Average SER}
		\includegraphics[width=42mm]{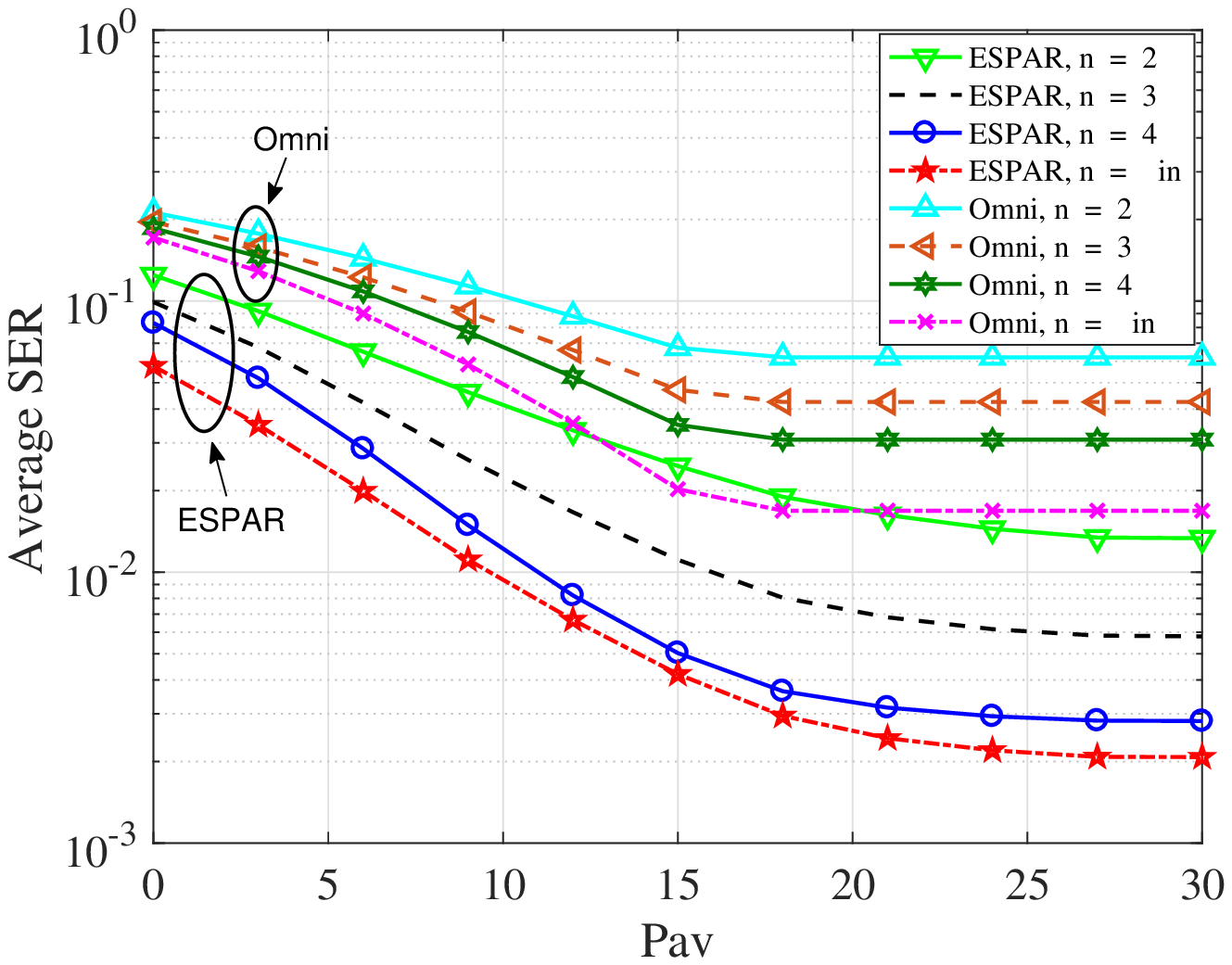}
		\vspace{-1mm}
		\caption{}
		\label{Pe_I6}
      \end{subfigure} \\
      \vspace{-1mm}
\caption{(a) $\overline{P}_\text{out}$ {\color{blue} and $P^\text{Om}_\text{out}$ }versus $\Pbar$, (b) $\overline{P}_\text{e}$ {\color{blue}and $P^\text{Om}_\text{e}$} versus $\Pbar$.}
\vspace{-4mm}
\end{figure}

 Let  $\overline{P}_\text{out}$ and $\overline{P}_\text{e}$ denote ${P_\text{out}}$ and ${P_\text{e}}$ that are  the averaged over all possible $\phi_\text{SR}^*$ and $\phi_\text{PU}^*$, respectively. {\color{blue} For comparison, we also include the outage and symbol error probabilities $P_\text{Out}^\text{Om}$ and $P_\text{e}^\text{Om}$ corresponding to the CR system that its \SUTx  ~has an omni-directional antenna.} 
Fig. \ref{Pout_I6} illustrates $\overline{P}_\text{out}$ {\color{blue} and $P_\text{out}^\text{Om}$} versus $\Pbar$. We observe that {\color{blue}  given an $n_b$ value, both outage and symbol error probabilities 
decrease as $\Pbar$ increases. However, they remain constant as  $\Pbar$ increases beyond a certain point (they reach error floors). These behaviors can be explained as the following.
{\color{blue}
For low $\Pbar$, the average transmit power constraint in \eqref{Pav01} is dominant and $\overline{P}_\text{out}$ and $\overline{P}_\text{e}$ decrease as $\Pbar$ increases, since \SUTx ~can transmit at higher power levels. On the other hand, for high  $\Pbar$, the average interference constraint in \eqref{Iav} is dominant and \SUTx ~cannot increase its transmit power level, regardless of how high $\Pbar$ becomes. As a result, $\overline{P}_\text{out}$ and $\overline{P}_\text{e}$ remain constant. 
Compared with the ESPAR antenna, the omni-directional antenna imposes a larger interference on PU. Thus, the average interference constraint for the omin-directional antenna becomes active at a smaller $\Pbar$ value, compared with the ESPAR antenna. As a result both outage and symbol error probabilities 
reach error floors at smaller $\Pbar$ values, compared with the ESPAR antenna.
Also, we note that as} $n_b$ increases $\overline{P}_\text{out}$ decreases. Fig. \ref{Pe_I6} plots $\overline{P}_\text{e}$ and $P^\text{Om}_\text{e}$ versus $\Pbar$.}  Similar observations to those of Fig. \ref{Pout_I6} can be made here. {\color{blue} In a nutshell, Figs. \ref{Pout_I6} and \ref{Pe_I6} show that  our proposed CR system yields lower outage and symbol error probabilities, 
compared with the CR system that its \SUTx  ~has an omni-directional antenna.} 
%
%
%
%
%
%
%
%
%
\section{Conclusions}\label{Conclusion}
%
%
We proposed a holistic system design for integrated sector-based spectrum sensing and sector-based data communication for an opportunistic CR system consisting of a PU, \SUTx, and \SURx, where \SUTx ~is equipped with an ESPAR antenna that has $M$ parasitic elements, and there is an error-free bandwidth limited feedback channel from \SURx ~to \SUTx. We formulated a constrained optimization problem, where the ergodic capacity for \SUTx-\SURx ~link is maximized, subject to average transmit power and interference constraints, and the optimization variables  are channel sensing duration, quantization thresholds at \SURx, and discrete power levels at \SUTx. Our problem formulation takes into consideration the effect of imperfect spectrum sensing, the error in determining the true orientation of PU, the error in selecting the strongest channel for data communication, and the impact of channel gain quantization. We developed an iterative suboptimal algorithm with a low computational complexity, based on the BCDA, that finds a unique and locally optimal solution for the constrained problem.  In addition, we derived closed form expressions for outage and symbol error probabilities of our opportunistic CR system. We corroborated our mathematical analyses with extensive simulations. Our {\color{blue} numerical} results demonstrate that our proposed CR system with the ESPAR antenna at \SUTx ~yields a significantly higher capacity, a lower outage probability, and a lower symbol error probability, compared with a CR system that its \SUTx ~has an omni-directional antenna. 
{\color{blue} The capacity improvement varies as the average transmit power and average interference constraints change. For instance, at $\Pbar\!=\!12$\,dB, $\Ibar\!=\!-6$\,dB, the capacity of our CR system is $1.83$ times larger than the capacity of the CR system with omni-directional antenna.} 
Furthermore, we showed that with only a small number of feedback bits the capacity of our CR system approaches to its baseline, which assumes the full knowledge of unquantized channel gain. 
%
%
%
%
\bibliographystyle{IEEEtran}
\bibliography{JournalRef}
%
\end{document}